\documentclass{aa}
\newif\iffigure
\figurefalse
\figuretrue

\newcommand{\edit}[1]{\textcolor{black}{#1}}
\newcommand{\editsec}[1]{\textcolor{black}{#1}}
\newcommand{\editthird}[1]{\textcolor{black}{#1}}
\newcommand{\editforth}[1]{\textcolor{black}{#1}}

\DeclareRobustCommand{\erase}{\bgroup\markoverwith{\textcolor{red}{\rule[.5ex]{2pt}{2pt}}}\ULon}

\usepackage[switch]{lineno}

\usepackage[dvipsnames]{xcolor}
\usepackage{graphicx,natbib,url,twoopt}
\usepackage[varg]{txfonts}           
\usepackage{hyperref}              
\usepackage{pdfcomment}              
\usepackage{acronym}                 
\usepackage{comment}
\usepackage{ulem}                    
\usepackage{cases}
\usepackage{empheq}                  
\usepackage{here}
\usepackage{bm}
\usepackage{tabularx}
\usepackage[version=3]{mhchem}
\hypersetup{
 colorlinks=true,%
 linkcolor=blue,
 citecolor=blue,
}

\usepackage{framed}
\usepackage{mdframed}

\newcommand*\patchAmsMathEnvironmentForLineno[1]{
  \expandafter\let\csname old#1\expandafter\endcsname\csname #1\endcsname
  \expandafter\let\csname oldend#1\expandafter\endcsname\csname end#1\endcsname
  \renewenvironment{#1}
     {\linenomath\csname old#1\endcsname}
     {\csname oldend#1\endcsname\endlinenomath}}
\newcommand*\patchBothAmsMathEnvironmentsForLineno[1]{
  \patchAmsMathEnvironmentForLineno{#1}
  \patchAmsMathEnvironmentForLineno{#1*}}
\AtBeginDocument{
\patchBothAmsMathEnvironmentsForLineno{equation}
\patchBothAmsMathEnvironmentsForLineno{align}
\patchBothAmsMathEnvironmentsForLineno{flalign}
\patchBothAmsMathEnvironmentsForLineno{alignat}
\patchBothAmsMathEnvironmentsForLineno{gather}
\patchBothAmsMathEnvironmentsForLineno{multline}
}

\bibpunct{(}{)}{;}{a}{}{,}    

\makeatletter
\newcommand{\bibnote}[2]{\global\@namedef{#1note}{#2}}
\newcommand{\biblink}[2]{\global\@namedef{#1link}{#2}}
\newcommand{\Tabref}[1]{Table~\ref{#1}}
\newcommand{\Equref}[1]{Eq.~\ref{#1}}
\newcommand{\Figref}[1]{Fig.~\ref{#1}}
\makeatother

\makeatletter
 \newcommandtwoopt{\citeads}[3][][]{%
   \nonstopmode
   \href{http://adsabs.harvard.edu/abs/#3}%
        {\def\hyper@linkstart##1##2{}%
         \let\hyper@linkend\@empty\citealp[#1][#2]{#3}}
   \biblink{#3}{\href{http://adsabs.harvard.edu/abs/#3}{ADS}}%
   \errorstopmode}            
 \newcommandtwoopt{\citepads}[3][][]{%
   \nonstopmode
   \href{http://adsabs.harvard.edu/abs/#3}%
        {\def\hyper@linkstart##1##2{}%
         \let\hyper@linkend\@empty\citep[#1][#2]{#3}}
   \biblink{#3}{\href{http://adsabs.harvard.edu/abs/#3}{ADS}}%
   \errorstopmode}            
 \newcommandtwoopt{\citetads}[3][][]{%
   \nonstopmode
   \href{http://adsabs.harvard.edu/abs/#3}%
        {\def\hyper@linkstart##1##2{}%
         \let\hyper@linkend\@empty\citet[#1][#2]{#3}}
   \biblink{#3}{\href{http://adsabs.harvard.edu/abs/#3}{ADS}}%
   \errorstopmode}            
 \newcommandtwoopt{\citeyearads}[3][][]{%
   \nonstopmode
   \href{http://adsabs.harvard.edu/abs/#3}%
        {\def\hyper@linkstart##1##2{}%
         \let\hyper@linkend\@empty\citeyear[#1][#2]{#3}}
   \biblink{#3}{\href{http://adsabs.harvard.edu/abs/#3}{ADS}}%
   \errorstopmode}            

\renewcommand{\@biblabel}[1]{[#1]} 

\makeatother

\newacro{ADS}{Astrophysics Data System}
\newacro{NLTE}{non-local thermodynamic equilibrium}
\newacro{NASA}{National Aeronautics and Space Administration}

\begin{document}
\authorrunning{Kuwahara and Lambrechts}
   \title{Interior dynamics of envelopes around disk-embedded planets}
      \author{Ayumu Kuwahara\inst{1} 
          \and Michiel Lambrechts\inst{1}}

   \institute{Center for Star and Planet Formation, GLOBE Institute, University of Copenhagen, Øster Voldgade 5-7, 1350 Copenhagen, Denmark
         }

   \date{Received September XXX; accepted YYY}

 
  \abstract
    {In the core accretion \editthird{scenario}, forming planets start to acquire gaseous envelopes \editsec{while accreting} solids. Conventional one-dimensional models assume envelopes to be static and isolated. \editthird{However, recent three-dimensional simulations demonstrate dynamic gas exchange from the envelope to the surrounding disk. This process is controlled by the balance between heating, through the accretion of solids, and cooling, which is regulated by poorly-known opacities.} In this work, we systemically investigate a wide range of cooling and heating rates, using three-dimensional hydrodynamical simulations. We identify three distinct cooling regimes. Fast-cooling envelopes \editsec{($\beta \lesssim 1$, with $\beta$ the cooling time in units of orbital time) are nearly isothermal and have inner} radiative layers that are shielded from recycling flows. \editsec{In contrast, slow cooling envelopes ($\beta\gtrsim10^3$)} become fully convective. \editsec{In the intermediate regime ($1\lesssim\beta\lesssim300$), envelopes are characterized by a three-layer structure, comprising an inner convective, a middle radiative, and an outer recycling layer.} \editsec{The development of this radiative layer traps small dust and vapour released from sublimated species. In contrast, fully convective envelopes efficiently exchange material from inner to outer envelope. Such fully convective envelopes are likely to emerge in the inner parts of protoplanetary disks ($\lesssim$ 1 au) where cooling times are long, implying that inner-disk super-Earths may see their growth stalled and be volatile depleted.}
    }

   \maketitle


\section{Introduction}\label{sec:Introduction}
In the framework of the core accretion model, planets form in protoplanetary disks \edit{where embryos first accrete planetesimals \citep{Pollack:1996} and pebbles \citep{Ormel:2010,Lambrechts:2012,Kuwahara:2020a,Kuwahara:2020b}}. Once the proto core exceeds approximately the \edit{mass of the Moon}, it begins to gravitationally capture gas from its surroundings, forming an envelope. As long as the planetary mass is below the so-called critical core mass of approximately 10 $M_\oplus$ (Earth masses), the envelope remains in \edit{near} hydrostatic equilibrium. Beyond this critical mass, the planets undergo runaway gas accretion, evolving into \edit{a} gas giant.

In classical one-dimensional (1D) \editsec{models}, the planetary envelopes are assumed to be \edit{static, isolated, and nearly spherically symmetric structures, embedded in the disk} \citep{Mizuno:1980,ikoma2000formation}. However, recent three-dimensional (3D) hydrodynamical simulations have revealed that envelopes are not closed systems, but rather interact dynamically with the surrounding disk through a process known as recycling \citep[e.g.,][]{Ormel:2015b,Fung:2015,Lambrechts:2017,Kuwahara:2019}.

\editsec{These recycling flows allow} for continuous gas exchange between the envelope and the disk, which could suppress the cooling and contraction \editsec{of the envelope}, thereby altering the timing and conditions under which runaway gas accretion occurs. \edit{Qualitatively, \editsec{envelope growth is} determined by the balance between heating and cooling. On the one hand, \editsec{planetary cores} grow by accreting solids, and the envelope heats up as the gravitational potential of the accreted solids is released. On the other hand, the envelope cools by radiation or convection, depending on its temperature gradient and ultimately on the opacity in the envelope \citep{rafikov2006atmospheres,piso2014minimum}.} Notably, \edit{previous works have shown that the} efficiency of this recycling process strongly depends on the radiative cooling efficiency of the gas \citep{Cimerman:2017,Kurokawa:2018,moldenhauer2021steady}.

The 3D dynamics of envelopes \edit{have further implications on the accretion of solids themselves. Gas flow around the planetary envelope} directly affect the motion of particles smaller than approximately millimeter in size (pebbles). Due to gas drag, the pebble accretion rate onto the planet is reduced by gas dynamics \citep{Popovas:2018a,Kuwahara:2020a,Kuwahara:2020b,okamura2021growth}. \edit{Moreover, as solids enter the higher temperature interior of the envelope, volatile species may sublimate and be advected along the gas flows through the envelope, thereby reducing the rate at which they are captured in the interior \citep{johansen2021pebble,wang2023atmospheric}.} 

\edit{Given these implications on gas and solids accretion rates onto embedded protoplanets,} it is important to investigate envelope dynamics, for a wide range of heating and cooling rates \edit{which has not been previously investigated}. Motivated by this, we present \editsec{in this study} a comprehensive investigation on the response of the envelope to a wide range of heating and cooling rates. Because cooling efficiency depends on the opacity within the envelope, which is subject to large uncertainties, we adopt a simplified model that parameterizes radiative cooling efficiency. 

\edit{The paper is structured as follows. Section \ref{sec:Numerical method} describes the numerical setup for our 3D hydrodynamical simulations that include accretion heating. In section \ref{sec:Numerical results},} we categorize the envelopes into three regimes, namely, fast, intermediate, and slow cooling regimes. We construct the analytic formulae of the mean flow speed within the envelope, then model the recycling efficiency of small particles and vapors in the envelope (Sect. \ref{sec:Tracing material transport}). We conclude in Sect. \ref{sec:Conclusions}.

\section{Numerical methods}\label{sec:Numerical method}
We simulate the heating and cooling processes of a pebble-accreting planet embedded in a non-self-gravitating disk with the Athena++ code\footnote{We used the public version of Athena++ (ver 24.0) which is available at: https://github.com/PrincetonUniversity/athena} \citep{stone2020athena++}. Our simulations were performed in spherical polar coordinates centered on a planet, where $r$ is the distance from the planet, $\theta$ the polar \editsec{angle}, and $\phi$ the azimuth angle. We use the default numerical settings of Athena++, such as the integration schemes, unless otherwise specified.

Assuming a compressible, inviscid, and non-self-gravitating fluid, the Athena++ code solves the equation of continuity, the Euler equation, and the energy equation:
\footnotesize
\editforth{\begin{align}
    &\frac{\partial \rho}{\partial t}+\nabla\cdot(\rho\bm{v})=0,\\
    &\frac{\partial (\rho\bm{v})}{\partial t}+\nabla\cdot(\rho \bm{v}\bm{v})=-\rho\nabla\Phi_{\rm p}-\nabla p+\rho(\bm{f}_{\rm cor}+\bm{f}_{\rm tid})\,,\\
    &\frac{\partial E}{\partial t}+\nabla\cdot\left[(E+p)\bm{v}\right]=-\rho\bm{v}\cdot\nabla\Phi_{\rm p}+\rho\bm{v}\cdot(\bm{f}_{\rm cor}+\bm{f}_{\rm tid})+Q_{\rm cool}+Q_{\rm acc}\,.\label{eq:energy eq}
\end{align}}
\normalsize
Here $\rho$ is the density, $\bm{v}$ is the velocity, and $p$ is the pressure. The total energy density is given by $E=e+\rho v^2/2$ and the internal energy density by $e=p/(\gamma-1)$ with $\gamma=1.4$ being the adiabatic index. \editforth{The Coriolis and tidal forces are given by $\bm{f}_{\rm cor}=-2\bm{e}_z\times\bm{v}_{\rm g}$ and $\bm{f}_{\rm tid}=3x\bm{e}_x-z\bm{e}_z$.} We include the thermal relaxation and accretion heating terms, $Q_{\rm cool}$ and $Q_{\rm acc}$, in the right-hand side of \Equref{eq:energy eq}, which is described later in Section \ref{sec:Cooling and heating terms}. 

We implemented the gravitational potential of the planet as
\begin{align}
    \Phi_{\rm p}=-\frac{GM_{\rm p}}{r}\times f_{\rm sm}\times f_{\rm inj},
\end{align}
where $G$ is the gravitational constant, and $M_{\rm p}$ is the planet mass. The smoothing function $f_{\rm sm}$ is given by \citep[]{Fung:2019}:
\begin{align}
    f_{\rm sm}=\frac{(r-r_{\rm in})^2}{(r-r_{\rm in})^2+r_{\rm sm}^2},\label{eq:Fung potential}
\end{align}
where $r_{\rm in}$ is the size of the inner boundary and $r_{\rm sm}$ is the smoothing length. We set $r_{\rm sm}=0.1\,r_{\rm in}$ \citep{zhu2021global}. \editsec{This formulation (Eq. \ref{eq:Fung potential}) gives} $\nabla\Phi_{\rm p}=0$ at the inner boundary, which is suitable for maintaining hydrostatic equilibrium. We note that a typical formula \editsec{for a smoothed} gravitational potential, $\Phi_{\rm p}=-GM_{\rm p}/\sqrt{r^2+r_{\rm sm}^2}$, leads to numerical heating, \editsec{because of} the non-zero gravitational force at the inner boundary (Appendix \ref{sec:Convergence tests}).

The gravity of the planet is gradually inserted into the disk to prevent shock formation. Following \cite{Ormel:2015a}\editsec{, we} used the following injection function 
\begin{align}
    f_{\rm inj}=1-\exp\Bigg[-\frac{1}{2}\Bigg(\frac{t}{t_{\rm inj}}\Bigg)^2\Bigg],
\end{align}
where $t$ is the time and $t_{\rm inj}=0.5\,\Omega^{-1}$ is the injection time and $\Omega$ is the orbital frequency.

\subsection{Cooling and heating terms}\label{sec:Cooling and heating terms}
We treat the cooling of the gas as a thermal relaxation process. This is implemented as a source term of the energy equation,
\begin{align}
    Q_{\rm cool}=-\frac{e-e_{\rm 0}}{t_{\rm cool}},
\end{align}
where $e_0$ represents the initial value. The isothermal and adiabatic limits are achieved when \edit{the cooling timescale approaches, respectively,} $t_{\rm cool}\rightarrow0$ and $t_{\rm cool}\rightarrow\infty$. 

We consider an envelope heated by pebble accretion with a luminosity \citep{rafikov2006atmospheres,Lambrechts:2014}: 
\begin{align}
    L(r)\approx\frac{GM_{\rm p}\dot{M}}{r_{\rm in}}\Bigg(1-\frac{r_{\rm in}}{r}\Bigg)\equiv L_{\rm acc}\Bigg(1-\frac{r_{\rm in}}{r}\Bigg). \label{eq:Lacc}
\end{align}
\editsec{Here, we consider envelope heating due the accretion of solids at a mass rate $\dot M$ and neglect} the other heat sources: the contraction of the envelope, the latent heat from evaporation, and the radioactive heating. The equation for the energy transportation is then given by
\begin{align}
    \frac{\mathrm{d}L(r)}{\mathrm{d}r}=4\pi r^2Q_{\rm acc}, \label{eq:dLacc dr}
\end{align}
where $Q_{\rm acc}$ is the heating rate per unit volume. From Eqs. (\ref{eq:Lacc}) and (\ref{eq:dLacc dr}), we have,
\begin{align}
    Q_{\rm acc}=\frac{L_{\rm acc}}{4\pi}\frac{r_{\rm in}}{r^4}.\label{eq:Zhu luminosity}
\end{align}
\editsec{This expression is also used by \cite{zhu2021global},} as it has the additional numerical benefit of no large intensity at the inner boundary. We confirmed that a jump appears in a temperature profile at the inner boundary when we consider the accretion heating only at the planet's surface (Appendix \ref{sec:Convergence tests}). The jump would lead to a large temperature gradient and hence to numerical convection.
\subsection{Code units and simulation parameters}

\begin{table*}[htbp]
\caption{Parameters of hydrodynamical simulations.}
\resizebox{\textwidth}{!}{
\begin{tabular}{lccccccc}\hline\hline
     & $m$ & $\beta$ & $\dot{M}$ & $N_r$ & \texttt{xorder} & $t_{\rm end}$ & $r_{\rm in}$ \\ \hline
     Fiducial runs & $0.1$ & 0.01, 0.1, 1, 10, 30, 100, 300, 1000, 10000 & $10\,M_\oplus$/Myr & 256 & 2 & 50 & $10\,R_{\rm p}/H$\\
     & $0.2$ & 10, 30, 100, 300, 1000 & $10\,M_\oplus$/Myr & 256 & 2 & 50 & $10\,R_{\rm p}/H$\\
     & $0.3$ & 10, 30, 100, 300, 1000 & $10\,M_\oplus$/Myr & 256 & 2 & 50 & $10\,R_{\rm p}/H$\\\hline
     Low luminosity runs & 0.1 & 10, 100, 1000 & $1\,M_\oplus$/Myr & 256 & 2 & 50 & $10\,R_{\rm p}/H$\\
      & 0.2 & 10, 100, 1000 & $1\,M_\oplus$/Myr & 256 & 2 & 50 & $10\,R_{\rm p}/H$\\
      & 0.3 & 10, 100, 1000 & $1\,M_\oplus$/Myr & 256 & 2 & 50 & $10\,R_{\rm p}/H$\\\hline
     High luminosity runs & 0.1 & 10, 100, 1000 & $100\,M_\oplus$/Myr & 256 & 2 & 50 & $10\,R_{\rm p}/H$\\
      & 0.2 & 10, 100, 1000 & $100\,M_\oplus$/Myr & 256 & 2 & 50 & $10\,R_{\rm p}/H$\\
      & 0.3 & 10, 100, 1000 & $100\,M_\oplus$/Myr & 256 & 2 & 50 & $10\,R_{\rm p}/H$\\\hline
     Convergence tests & $0.1$ & 1, 1000 & $10\,M_\oplus$/Myr & 168 & 2, 3 & 10 & $10\,R_{\rm p}/H$\\    
     & $0.1$ & 1, 1000 & $10\,M_\oplus$/Myr & 196 & 2, 3 & 10 & $10\,R_{\rm p}/H$\\    
     & $0.1$ & 1, 10, 100, 1000 & $10\,M_\oplus$/Myr & 256 & 3 & 10 & $10\,R_{\rm p}/H$\\    
     & $0.1$ & 1, 1000 & $10\,M_\oplus$/Myr & 256 & 2 & 10 & $5\,R_{\rm p}/H$\\
     \hline
\end{tabular}
}
\tablefoot{The following columns give the dimensionless thermal mass, the dimensionless cooling time, the pebble accretion rate, the radial resolution, the method for spatial reconstruction, the calculation time, and the size of the inner boundary.}
\label{tab:hydro simulations}
\end{table*}

Our simulations were performed in the units of $H=c_{\rm s}=\Omega=\rho_0=1$, where $H$ is the scale height, $c_{\rm s}$ is the isothermal sound speed, and $\rho_0$ is the midplane gas density at the planet orbital location, $a_{\rm p}$. Since we neglect the self-gravity of the disk gas, we can introduce another normalization for the planetary mass,
\begin{align}
    m\equiv\frac{R_{\rm B}}{H}&=\frac{M_{\rm p}}{M_{\rm th}}\simeq0.11\,\Bigg(\frac{M_{\rm p}}{M_\oplus}\Bigg)\Bigg(\frac{M_\ast}{M_\odot}\Bigg)^{-1}\Bigg(\frac{0.03}{h}\Bigg)^3\editsec{.}
\end{align}
\editsec{Here,} \edit{$R_{\rm B}=GM_{\rm p}/c_{\rm s}^2$ is the Bondi radius}, $M_{\rm th}=M_\ast h^3$ is the thermal mass, $M_\ast$ is the stellar mass, $h$ is the disk aspect ratio, \editsec{$M_\odot$ is the solar mass. Finally,} we parameterized the thermal relaxation timescale as the dimensionless quantity \citep{Gammie:2001},
\begin{align}
    \beta\equiv t_{\rm cool}\Omega.
\end{align}

\editsec{In our simulation, we explore different solid accretion rates. The accretion luminosity (Eq. \ref{eq:Lacc}) at the planetary surface $r=R_{\rm p}$ is of the order}
\begin{align}
    L_{\rm surf}=\frac{GM_{\rm p}\dot{M}}{R_{\rm p}}\sim10^{27}\,\text{erg/s}\,\Bigg(\frac{M_{\rm p}}{M_{\oplus}}\Bigg)^{2/3}\Bigg(\frac{\dot{M}}{10\,M_{\oplus}/\text{Myr}}\Bigg).
\end{align} 
The dimensionless accretion luminosity is then given by
\footnotesize
\begin{align}
    \tilde{L}_{\rm acc}&=\frac{L_{\rm surf}}{c_{\rm s}^2M_{\rm th}\Omega}\nonumber\\
    &\simeq1.1\times10^{-5}\,\Bigg(\frac{m}{0.1}\Bigg)^{5/3}\Bigg(\frac{\rho_{\rm p}}{5\,\text{g/cm}^{3}}\Bigg)^{1/3}\Bigg(\frac{M_\ast}{M_\odot}\Bigg)^{-5/6}\Bigg(\frac{a_{\rm p}}{1\,\text{au}}\Bigg)^{5/2}\Bigg(\frac{t_{\rm acc}}{10^5\,\text{yr}}\Bigg)^{-1},
\end{align}
\normalsize
where we used,
\begin{align}
    \frac{R_{\rm p}}{H}\simeq1.4\times10^{-3}\,\Bigg(\frac{m}{0.1}\Bigg)^{1/3}\Bigg(\frac{\rho_{\rm p}}{5\,\text{g/cm}^3}\Bigg)^{-1/3}\Bigg(\frac{M_\ast}{M_\odot}\Bigg)^{1/3}\Bigg(\frac{a_{\rm p}}{1\,\text{au}}\Bigg)^{-1}.
\end{align}
Here $\rho_{\rm p}$ is the density of the planet. We defined the accretion timescale as \editsec{$t_{\rm acc}\equiv M_{\rm p}/\dot{M}$.}
For fiducial runs, we set $t_{\rm acc}=10^5$ yr, which corresponds to pebble accretion rates of $\dot{M}=10\,M_\oplus/$Myr \citep{Lambrechts:2014}. We also investigate low- and high-luminosity cases with $t_{\rm acc}=10^6$ yr and $10^4$ yr (corresponding to $\dot{M}=1\,M_\oplus/$Myr and $\dot{M}=100\,M_\oplus/$Myr), \editsec{where the latter case explores the upperend of possible solid accretion rates}. The parameters of our simulations are summarized in Table \ref{tab:hydro simulations}.


\subsection{Initial and boundary conditions, and resolutions}
We assume a vertically stratified density profile \editsec{for} the initial condition, 
\begin{align}
    \frac{\rho}{\rho_0}=\exp\Bigg[-\frac{1}{2}\bigg(\frac{z}{H}\bigg)^2\Bigg].
\end{align}
\editsec{A} Keplerian shear flow \editsec{is applied} as an initial background velocity field, neglecting the headwind of the gas due to a global pressure gradient in a disk \editsec{(see also Sect. \ref{sec:Model limitation: pure Keplerian rotation})}. The initial velocity is given by
\begin{align}
    \frac{\bm{v}}{c_{\rm s}}=-\frac{3}{2}\frac{x}{H}\,\bm{e}_y.
\end{align}

We use a log-spaced grid in the radial coordinate ranging from $r_{\rm in}=10\,R_{\rm p}/H$ to $r_{\rm out}=10\,R_{\rm B}/H$, which is divided into 256 grids. \editsec{The polar and azimuth angles are uniformly divided into 40 and 160 grids.} 

\editsec{T}he logarithmic grids have a higher resolution at the region close to the planet\editsec{, so that the} Bondi radius is resolved by at least 118 grids in the radial direction. 
To save the computational costs, the size of the inner boundary was arbitrarily determined \editsec{to be $r_{\rm in}=10\,R_{\rm p}$}. We confirmed that the results hardly change when we adopted $r_{\rm in}=5\,R_{\rm p}$ (Appendix \ref{sec:Convergence tests}). 

\editsec{F}or spatial reconstruction, we use a second-order piecewise linear method (PLM; \texttt{xorder=2} in the Athena++). In Appendix \ref{sec:Convergence tests}, we show the agreement between simulations using the PLM method and those using a higher-order method (a fourth-order piece wise parabolic method, PPM; \texttt{xorder=3}).  

For the radial direction we set \editsec{an} inner reflecting boundary condition. At the outer boundary, we fix the density and the velocity \editsec{to} the initial values. We only consider the upper half region of the disk, $\theta\in[0,\,\pi/2]$. We use \editsec{a} reflecting condition at the midplane, $\theta=0$. On the pole we use the polar boundary condition, in which the physical quantities in the ghost cells are copied from the other side of the pole \citep{stone2020athena++}. We consider the full range of the azimuthal angle, $\phi\in[0,\,2\pi]$.



\begin{figure}[t]
    \centering
    \includegraphics[width=0.97\linewidth]{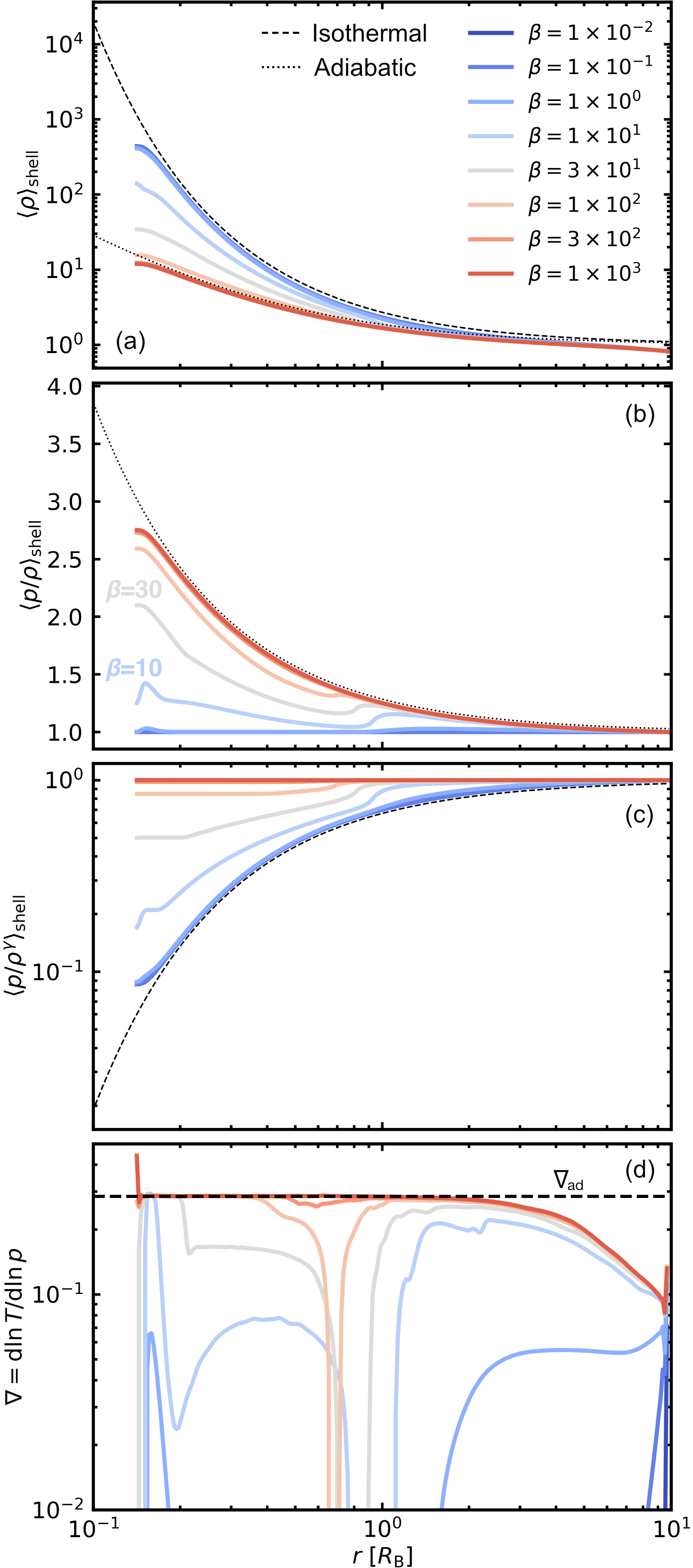}
    \caption{Shell-averaged density, temperature, entropy, and temperature gradient (from top to bottom) for different $\beta$ values. We set $m=0.1$ and $\dot{M}=10\,M_\oplus/$Myr. All panels are the snapshots at the end of the calculation, $t=50$. The dashed and dotted curves in panels a--c are the analytic models given by Eqs. (\ref{eq:iso den profile}), (\ref{eq:ad den profile}), and (\ref{eq:ad temp profile}). The horizontal dashed line in panel d is the adiabatic gradient, \Equref{eq:Schwarzschild criterion}.}
    \label{fig:den_temp_entropy}
\end{figure}

\begin{figure}
    \centering
    \includegraphics[width=1\linewidth]{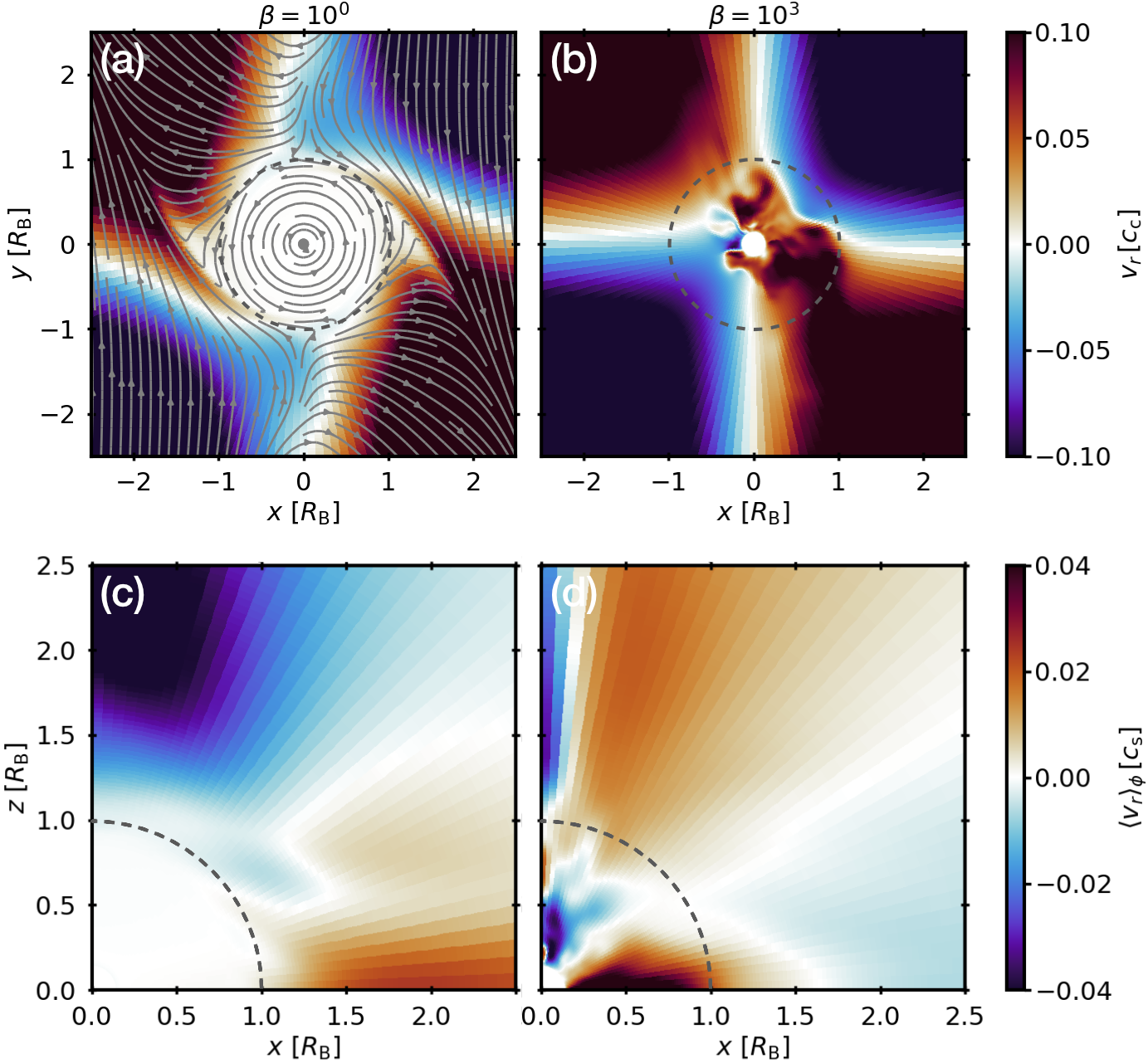}
    \caption{Radial velocity of the gas\editsec{,} as a function of \editsec{cooling time} $\beta$. We set $m=0.1$ and $\dot{M}=10\,M_\oplus/$Myr. All panels are snapshots at the end of the calculation, $t=50$. The dashed circle denotes the Bondi radius. \textit{Top}: \editsec{Midplane slices, with streamlines shown} in \editforth{panel a}. \textit{Bottom}: \editsec{Vertical slices where the velocity is} hemispherically averaged in the azimuthal direction, $\phi\in[-\pi/2,\pi/2]$}
    \label{fig:vr}
\end{figure}

\begin{figure*}
    \centering
    \includegraphics[width=1\linewidth]{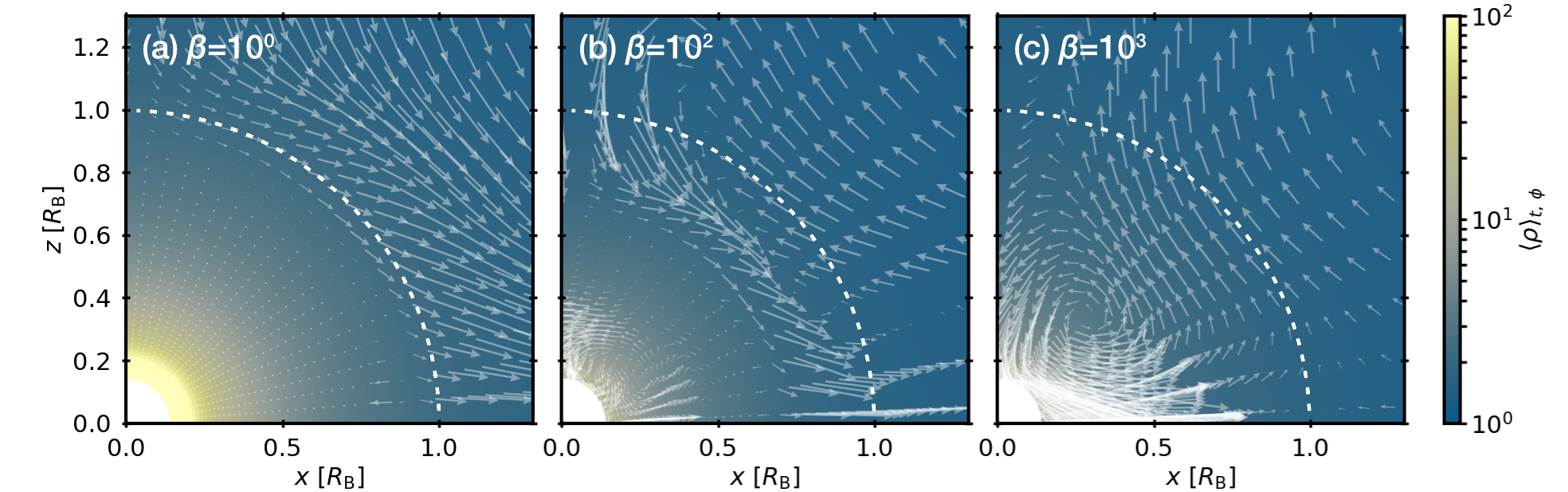}
    \caption{Hemispherically- and time-averaged density and velocity vector on the vertical slice. We set $m=0.1$. The physical quantities are averaged over the azimuth and the time, $\phi\in[-\pi/2,\pi/2]$ and $t\in[40,50]$. The white dash circle in the top panels is the Bondi radius of the planet.}
    \label{fig:azimuth_and_time_average}
\end{figure*}

\begin{figure}[t]
    \centering
    \includegraphics[width=1\linewidth]{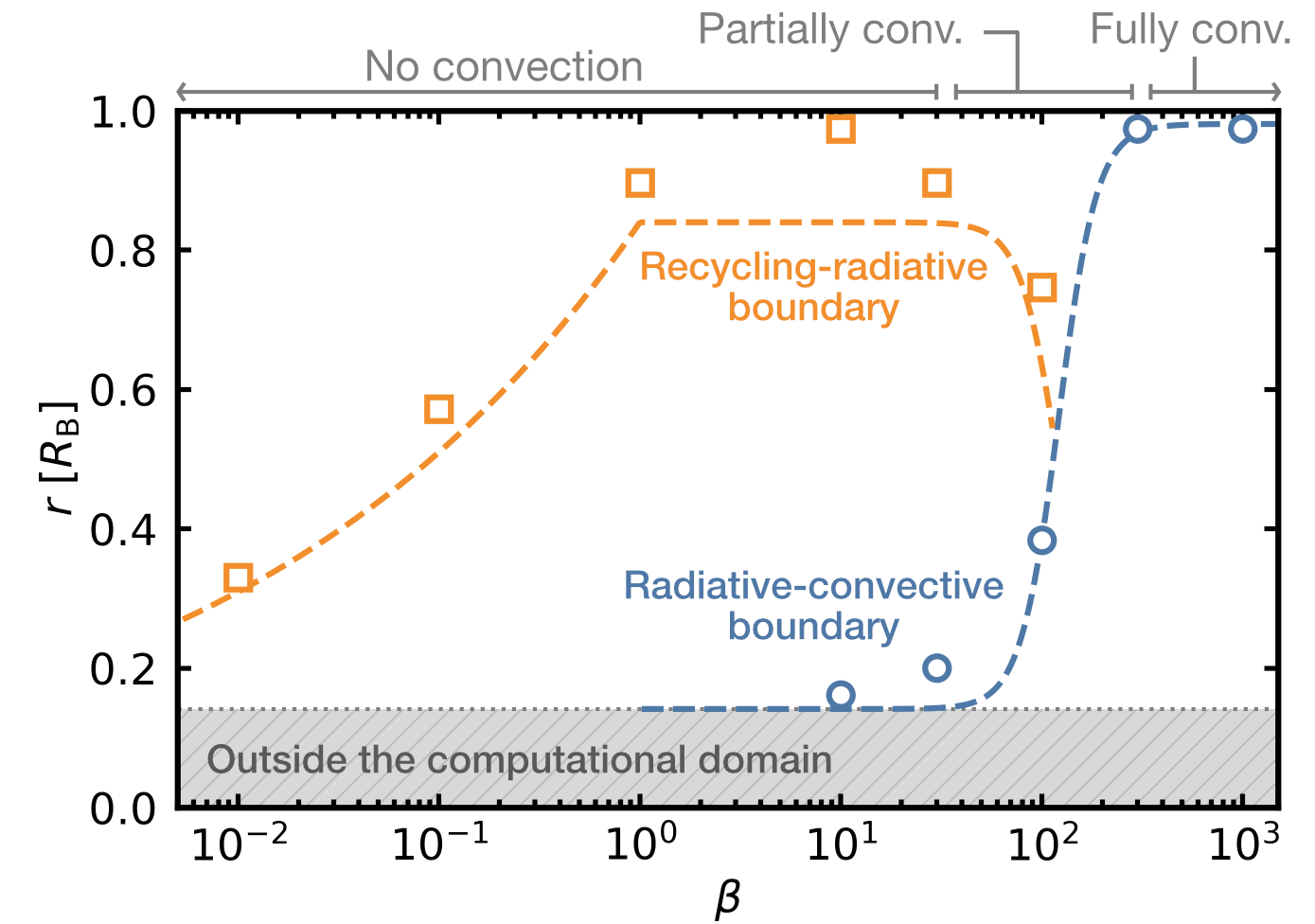}
    \caption{\editsec{Boundary between the recycling and radiative layers (orange) and the radiative and convective layers (blue), as a function of cooling time $\beta$. The squares and the circles are the numerical results\editthird{, obtained for $m=0.1$ and $\dot{M}=10\,M_\oplus/$Myr}. The dashed lines are the fitting formulae (Appendix \ref{sec:Appendix fitting formulae}). The hatched region is not probed by our simulations.}}
    \label{fig:envelope_layer}
\end{figure}

\begin{figure}
    \centering
    \includegraphics[width=1\linewidth]{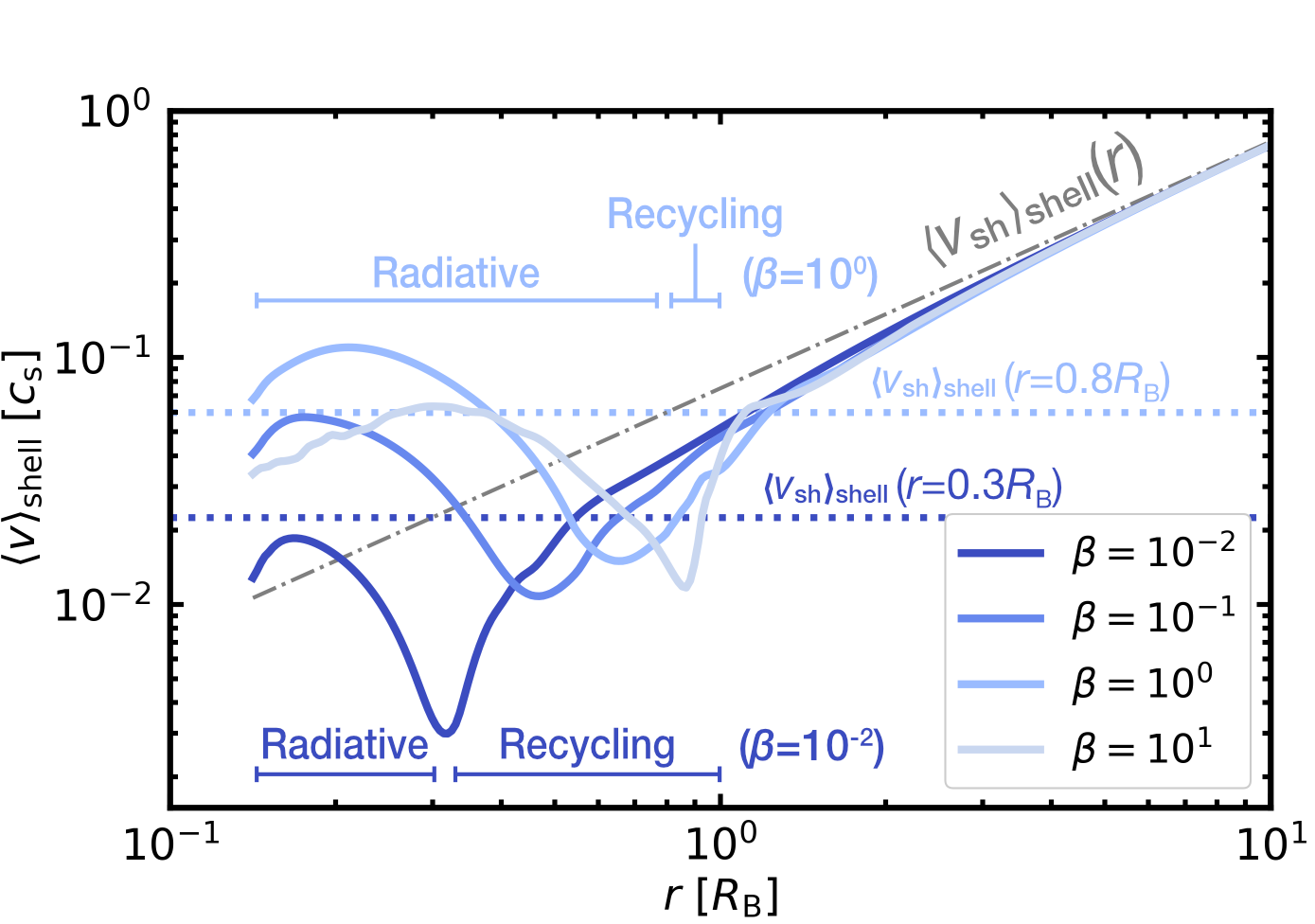}
    \caption{Shell-averaged mean speed for $\beta=10^{-2}\text{--}10^1$. We set $m=0.1$ and $\dot{M}=10\,M_\oplus/$Myr. This panel shows \editsec{the state} at the end of the simulation, $t=50$. The radiative and recycling layers are qualitatively indicated. The horizontal dotted lines correspond to \Equref{eq:vsh_shell} with $r=0.3\,R_{\rm B}$ and \editsec{$r=0.8\,R_{\rm B}$}. The gray dashed-dotted curve represents \Equref{eq:vsh_shell}.}
    \label{fig:shell_avg_vel_b1e-2_1e1}
\end{figure}

\section{Numerical results}\label{sec:Numerical results}

\editsec{We find that the} (thermo)dynamic states of planetary envelopes 
can be classified into three categories: \editsec{the} fast cooling ($\beta<1$) \editsec{regime, the} intermediate cooling ($1\lesssim\beta<1000$) \editsec{regime}, and the slow cooling ($\beta\gtrsim1000$) regime. The following sections (Sects. \ref{sec:Fast cooling regime}--\ref{sec:Slow cooling regime}) describe the properties of the envelope in each regime for a fiducial run with $m=0.1$ and $\tilde{L}_{\rm acc}=1.1\times10^{-5}\,(m/0.1)^{5/3}(t_{\rm acc}/10^5\,\text{yr})^{-1}$. The dependence \edit{on} the planetary mass and the accretion luminosity will be presented in Sect. \ref{sec:Planetary mass and luminosity dependence}.

\subsection{Fast cooling regime}\label{sec:Fast cooling regime}
When $\beta\lesssim1$, an envelope is nearly isothermal\editsec{,} \edit{corresponding to} efficient radiative cooling. \edit{The shell-averaged density, temperature, and entropy profiles are in good agreement with the isothermal hydrostatic solution (\Figref{fig:den_temp_entropy}). Here, the hydrostatic solution of the density is given by \citep[e.g.,][]{rafikov2006atmospheres}}
\begin{align}
    \rho_{\rm iso}(r)\approx\rho_0\exp\Bigg(\frac{R_{\rm B}}{\sqrt{r^2+r_{\rm sm}^2}}\Bigg).\label{eq:iso den profile}
\end{align}
\edit{Equation \ref{eq:iso den profile} is plotted in \Figref{fig:den_temp_entropy}a with the dashed line. In \Figref{fig:den_temp_entropy}b, we show the shell-averaged value of $p/\rho$ which is the measure of the temperature normalized by the value of the background disk, $T_0$. In this study, we define the entropy function as} 
\begin{align}
    \mathcal{S}\equiv\frac{p}{\rho^\gamma}.
\end{align}
\edit{The dashed line in \Figref{fig:den_temp_entropy}c corresponds to $p_{\rm iso}/\rho_{\rm iso}^\gamma$, with $p_{\rm iso}\equiv\rho_{\rm iso}c_{\rm s}^2$ being the pressure of an isothermal envelope.}

\editsec{T}he envelope is convectively stable when the temperature gradient falls below the adiabatic gradient,
\begin{align}
        \frac{\partial \ln T}{\partial \ln p}<\nabla_{\rm ad}\equiv\frac{\gamma-1}{\gamma}.\label{eq:Schwarzschild criterion}
\end{align}
\edit{In the fast cooling regime,} the temperature gradient is much smaller than the adiabatic gradient throughout the Bondi sphere (\Figref{fig:den_temp_entropy}d).

The outer envelope, typically outside of $0.5\,R_{\rm B}$, is subject to a recycling flow. Figure \ref{fig:vr} shows the radial velocity of the gas and its dependence on $\beta$ in both \edit{the $x$-$y$ midplane and the $x$-$z$ perpendicular plane}. \edit{The gas, on horseshoe-like orbits, enters the Bondi sphere from the polar regions and} exits through the midplane. \edit{This is} referred to as the recycling flow \citep[e.g.,][]{Ormel:2015b,Fung:2015,Kuwahara:2019}. However, the inner envelope remains bound to the planet with polar inflow unable to penetrate beyond a certain depth \editforth{(Fig.~\ref{fig:vr}, panel a and c)}. 
The isolation of the inner radiative layer from the recycling layer is clearly seen in the velocity vector shown in \Figref{fig:azimuth_and_time_average}a. \edit{This aligns with previous simulations, without including accretion heating,} that a positive entropy gradient (buoyancy force; \edit{\Figref{fig:den_temp_entropy}c}) suppresses the inflow penetrating the envelope \citep{Kurokawa:2018}.

\editsec{
\editforth{The boundary between the inner radiative and outer recycling layers depends on the cooling time $\beta$, and is typically located at approximately $0.5\text{--}1\,R_{\rm B}$, shifting outward with increasing $\beta$. Figure~\ref{fig:envelope_layer} shows the numerically-calculated recycling-radiative boundary (RRB) defined as,} 
\begin{align}
        r_{\rm RRB}^{\rm num}\equiv\max\Bigg\{r\in[0,R_{\rm B}]\,\bigg|&\,\langle v_{\phi,{\rm mid}}\rangle_\phi>\max\bigg(\langle v_{r,{\rm mid}}\rangle_\phi,\,\langle v_{\theta,{\rm mid}}\rangle_\phi\bigg)\nonumber\\
            &\land\,\nabla_{\rm shell}<0.95\nabla_{\rm ad}\Bigg\}.\label{eq:rrb num}
\end{align}
Here $\langle v_{i,{\rm mid}}\rangle_\phi$ is the azimuthally-averaged velocity at the midplane and $\nabla_{\rm shell}$ is the shell-averaged value of $\partial\ln T/\partial\ln p$. The first condition on the right-hand side of \Equref{eq:rrb num} corresponds to the dynamical definition of a bound atmosphere \citep{moldenhauer2022recycling}. The factor of 0.95 in the second condition is a \editthird{safety} factor \editthird{for the numerically determined temperature gradient $\nabla_{\rm shell}$ (\Figref{fig:den_temp_entropy}d)}. The orange squares in \Figref{fig:envelope_layer} mark $r_{\rm RRB}^{\rm num}$ for the fiducial run with $m=0.1$. A fitting formula for $r_{\rm RRB}^{\rm num}$ is given in Appendix \ref{sec:Appendix fitting formulae} (orange dashed line in \Figref{fig:envelope_layer}).
}

\edit{The mean flow speed in the radiative layer is characterized by the shear flow at its outer edge and reaches approximately 1 to 10\% of the sound speed, depending on $\beta$ (\Figref{fig:shell_avg_vel_b1e-2_1e1}). \editsec{Further outside} the envelope, $r\gg R_{\rm B}$, the flow is dominated by the Keplerian shear of which \editforth{hemispherically}-averaged value is \editsec{given} by}
\editforth{\begin{align}
    \displaystyle
    \frac{{\langle v_{\rm sh}\rangle_{\rm shell}}(r)}{c_{\rm s}} = \frac{\int_0^{2\pi}\int_0^{\pi} \frac{3}{2}r^3\Omega\sin^2\theta|\cos\phi|\mathrm{d}\theta\mathrm{d}\phi}{\int_0^{2\pi}\int_0^{\pi} r^2\sin\theta\mathrm{d}\theta\mathrm{d}\phi}\frac{1}{c_{\rm s}}=\frac{3}{4}\frac{r}{H}.\label{eq:vsh_shell}
\end{align}} In \Figref{fig:shell_avg_vel_b1e-2_1e1}, the predicted mean flow speeds from \Equref{eq:vsh_shell} are shown by the dashed lines for $\beta=0.01$ and 1, assuming that the outer edges of the radiative layer are $r=0.3\,R_{\rm B}$ and \editsec{$0.8\,R_{\rm B}$}, respectively. These predictions are consistent with the numerical results. The mean flow speed is lowest at the boundary between the radiative and recycling layers and converges to the Keplerian shear beyond the Bondi radius (\Figref{fig:shell_avg_vel_b1e-2_1e1}).

\begin{figure}[t]
    \centering
    \includegraphics[width=1\linewidth]{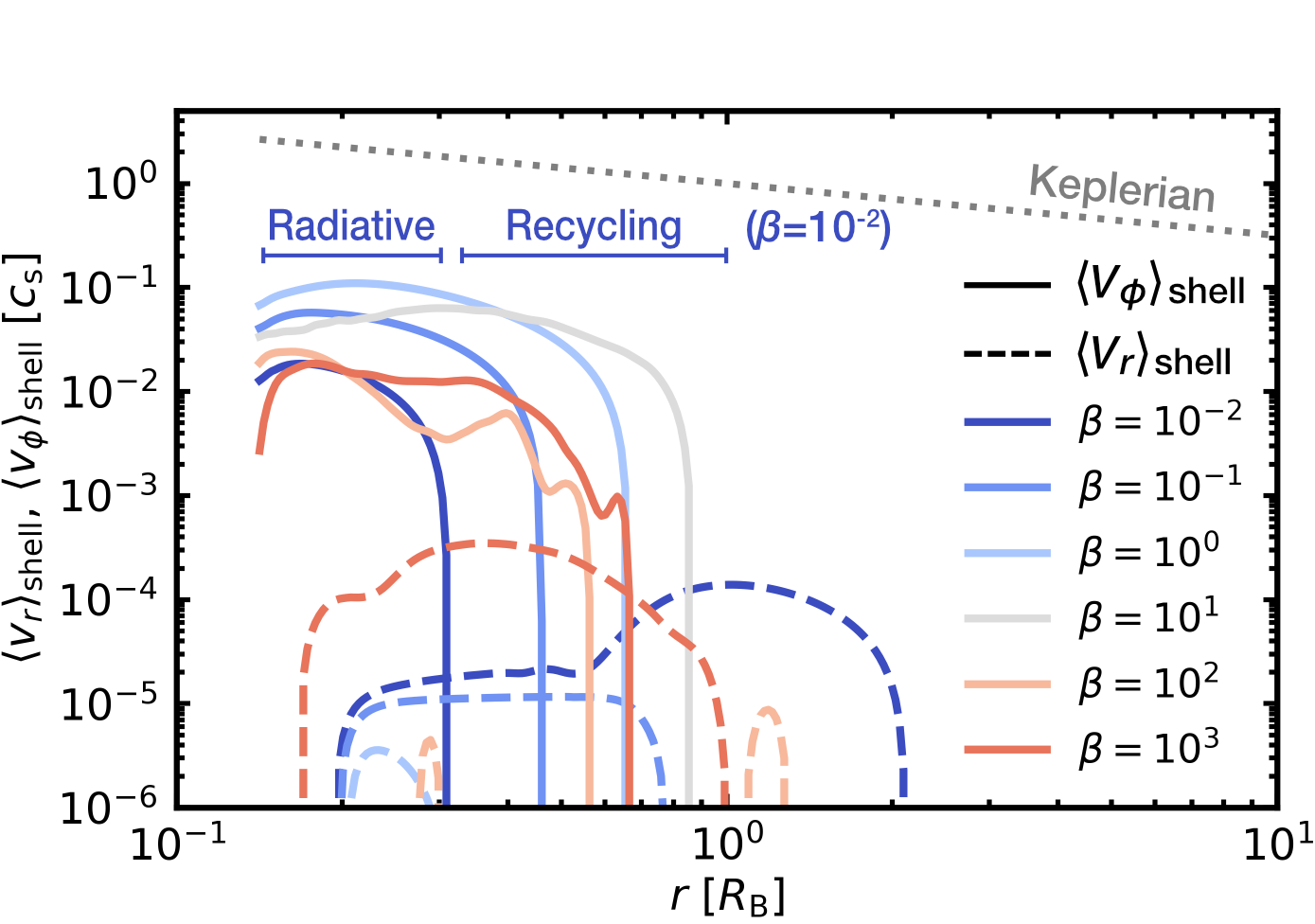}
    \caption{Shell-averaged radial \editsec{velocity (solid lines) and azimuthal velocity (dashed lines), for different $\beta$.} Negative components are omitted. We set $m=0.1$ and $\dot{M}=10\,M_\oplus/$Myr. The gray dotted line represents the Keplerian \editsec{velocity around the core}.}
    \label{fig:shell_avg_vel_vr_vphi}
\end{figure}

\edit{We show the shell-averaged radial and azimuthal velocities in \Figref{fig:shell_avg_vel_vr_vphi}.} The envelopes rotate in the prograde direction, opposite the direction of the shear flow. \editsec{We do not see the development of a  circumplanetary disk: the rotational velocity of the envelope is much smaller than the Keplerian velocity.} The formation of a prograde-rotating envelope can be attributed to the influence of the Coriolis force on the 3D gas flow \citep{Ormel:2015b}. 
The azimuthal velocity dominates in the radiative layer (the solid lines in \Figref{fig:shell_avg_vel_vr_vphi}), peaking near the inner boundary and decreasing outward.

\begin{figure}[t]
    \centering
    \includegraphics[width=1\linewidth]{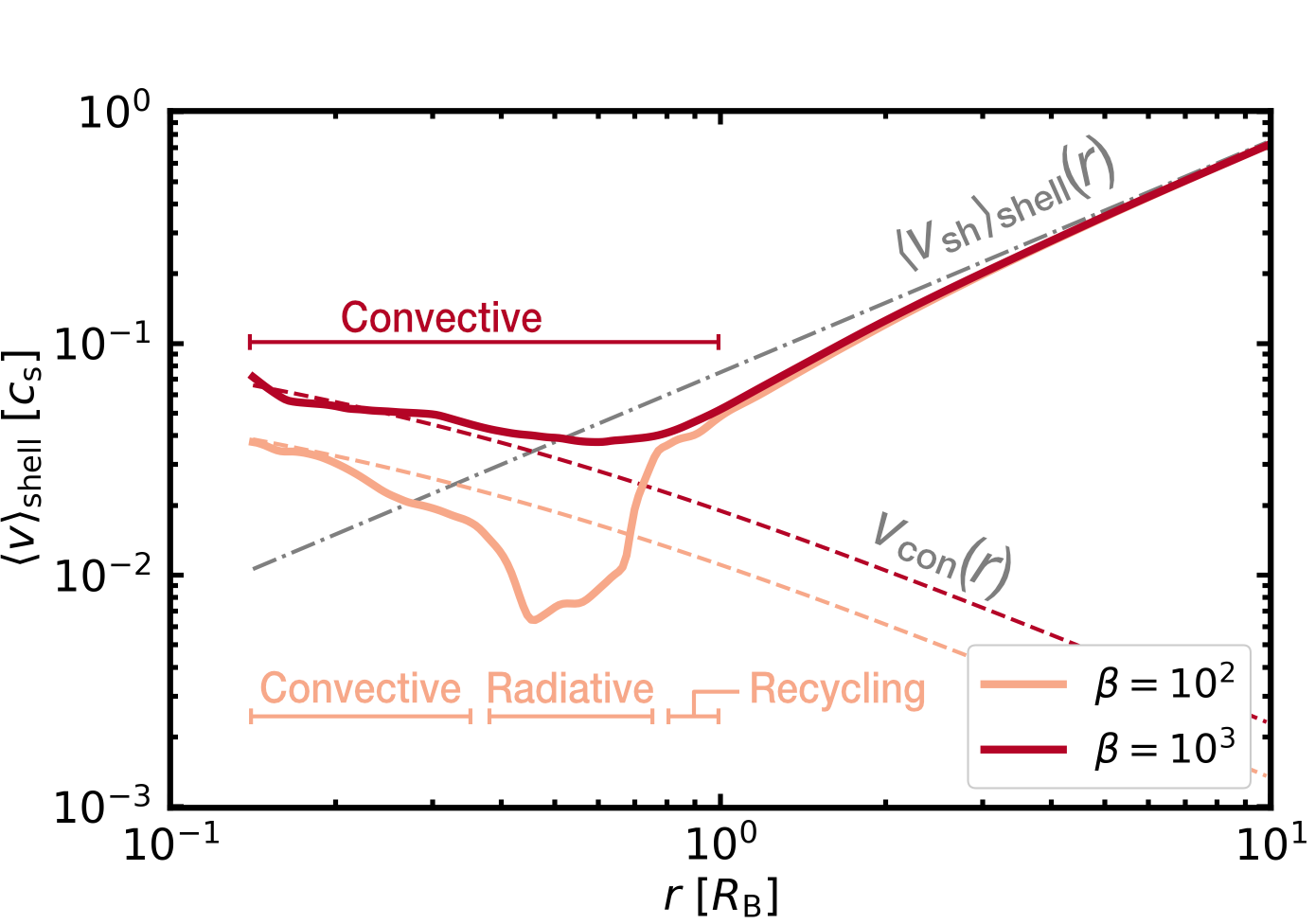}
    \caption{Same as Fig.~\ref{fig:shell_avg_vel_b1e-2_1e1}, but for $\beta=10^{2}$ and $10^3$. The convective, radiative, and recycling layers are qualitatively indicated. The red and yellow dashed curves show \Equref{eq:vcon}\editsec{,} under the assumptions of fully and partially convective envelopes \editsec{with} $l_{\rm m}=R_{\rm B}$ and $l_{\rm m}=0.2\,R_{\rm B}$, respectively.}
    \label{fig:shell_avg_vel_b1e2_1e3}
\end{figure}

\subsection{Intermediate cooling regime}\label{sec:Intermediate cooling regime}
This regime ($1\lesssim\beta\lesssim300$) features a three-layer envelope structure \edit{with, from inside to outside, a convectve, radiative and recycling layer \citep[\Figref{fig:azimuth_and_time_average}b;][]{Lambrechts:2017}.} The \edit{shell-averaged} envelope properties---temperature, density, and entropy---lie between isothermal and adiabatic profiles (Figs. \ref{fig:den_temp_entropy}a--c). As $\beta$ increases, the entropy gradient flattens (\Figref{fig:den_temp_entropy}c), allowing polar inflow to penetrate deeper (Figs. \ref{fig:azimuth_and_time_average}a and b).


\edit{Similar to the near-isothemal envelopes, the outer shells are dominated by Keplerian shear penetrating the envelope. However, deeper inside the envelope we notice convection motions bound to the inner envelope\editsec{,} when $30\lesssim\beta\lesssim300$ (\Figref{fig:azimuth_and_time_average}b).
\editsec{The blue circles in \Figref{fig:envelope_layer} mark the numerically-calculated radiative-convective boundary (RCB), 
\begin{align}
     r_{\rm RCB}^{\rm num}\equiv\max\Bigg\{r\in[0,R_{\rm B}]\,\bigg|\,\nabla_{\rm shell}\geq0.95\nabla_{\rm ad}\Bigg\},\label{eq:rcb num}
\end{align}
while the blue dashed line shows the corresponding fitting formula (see Appendix \ref{sec:Appendix fitting formulae}).} In Appendix \ref{sec:Characteristic convective flow speed}, we present an analytic model to calculate the shell-averaged mean flow speed in the convective layer, which is in good agreement with the numerical results (\Figref{fig:shell_avg_vel_b1e2_1e3}). \edit{When the energy is transported by convection, in the framework of mixing length theory, the characteristic speed of convection is given by (Appendix \ref{sec:Characteristic convective flow speed}):}
\begin{align}
    \frac{v_{\rm con}(r,l_{\rm m},\tilde{L}_{\rm acc})}{c_{\rm s}}\approx\Bigg(\frac{\nabla_{\rm ad}}{8\pi (r/H)^3}\frac{(l_{\rm m}/R_{\rm B})\tilde{L}_{\rm acc}}{\rho_{\rm ad}(r)/\rho_0}\Bigg)^{1/3}.\label{eq:vcon}
\end{align}
\editsec{Here,} $l_{\rm m}$ is the mixing length corresponding to the size of the convective cell. The convective flow speed is fastest near the inner boundary and decreasing outward (Eq. \ref{eq:vcon}; \Figref{fig:shell_avg_vel_b1e2_1e3}).} The maximum convective flow speed reaches a few percent of the sound speed. The mean flow speed drops by an order of magnitude in the radiative layer, then increases in the recycling layer due to the subsonic polar inflow and midplane outflow.

\subsection{Slow cooling regime}\label{sec:Slow cooling regime}
When $\beta\gtrsim1000$, the envelope is fully convective \editforth{(Fig.~\ref{fig:vr}, panel b and d)}. \edit{No steady-state flow can be identified within the envelope and material isotropically enters and leaves the envelope (\Figref{fig:azimuth_and_time_average}c). Nevertheless, the shell-averaged density, temperature, and entropy (Figs. \ref{fig:den_temp_entropy}a--c) match the 1D analytic adiabatic limit \editsec{\citep[e.g.,][]{rafikov2006atmospheres}:}} 
\begin{align}
    &\rho_{\rm ad}(r)\approx\rho_0\Bigg(1+\nabla_{\rm ad}\frac{R_{\rm B}}{\sqrt{r^2+r_{\rm sm}^2}}\Bigg)^{1/(\gamma-1)},\label{eq:ad den profile}\\
    &T_{\rm ad}(r)\approx T_0\Bigg(1+\nabla_{\rm ad}\frac{R_{\rm B}}{r}\Bigg).\label{eq:ad temp profile}
\end{align} 
The temperature gradient matches the adiabatic one throughout the Bondi sphere (\Figref{fig:den_temp_entropy}d).

\edit{Flow speeds in the convective envelope now reach approximately
10\% of the sound speed (\Figref{fig:shell_avg_vel_b1e2_1e3}). This is in agreement with \Equref{eq:vcon}}. Larger mixing lengths and lower densities compared to the intermediate regime lead to higher convective velocities, as predicted by \Equref{eq:vcon}, $v_{\rm con}\propto(l_{\rm m}L/\rho)^{1/3}$. 
This vigorous convection influences material transport, which is explored further in Sect. \ref{sec:Tracing material transport}.

\subsection{Planetary mass and luminosity dependence}\label{sec:Planetary mass and luminosity dependence}
\begin{figure}[tp]
    \centering
    \includegraphics[width=1\linewidth]{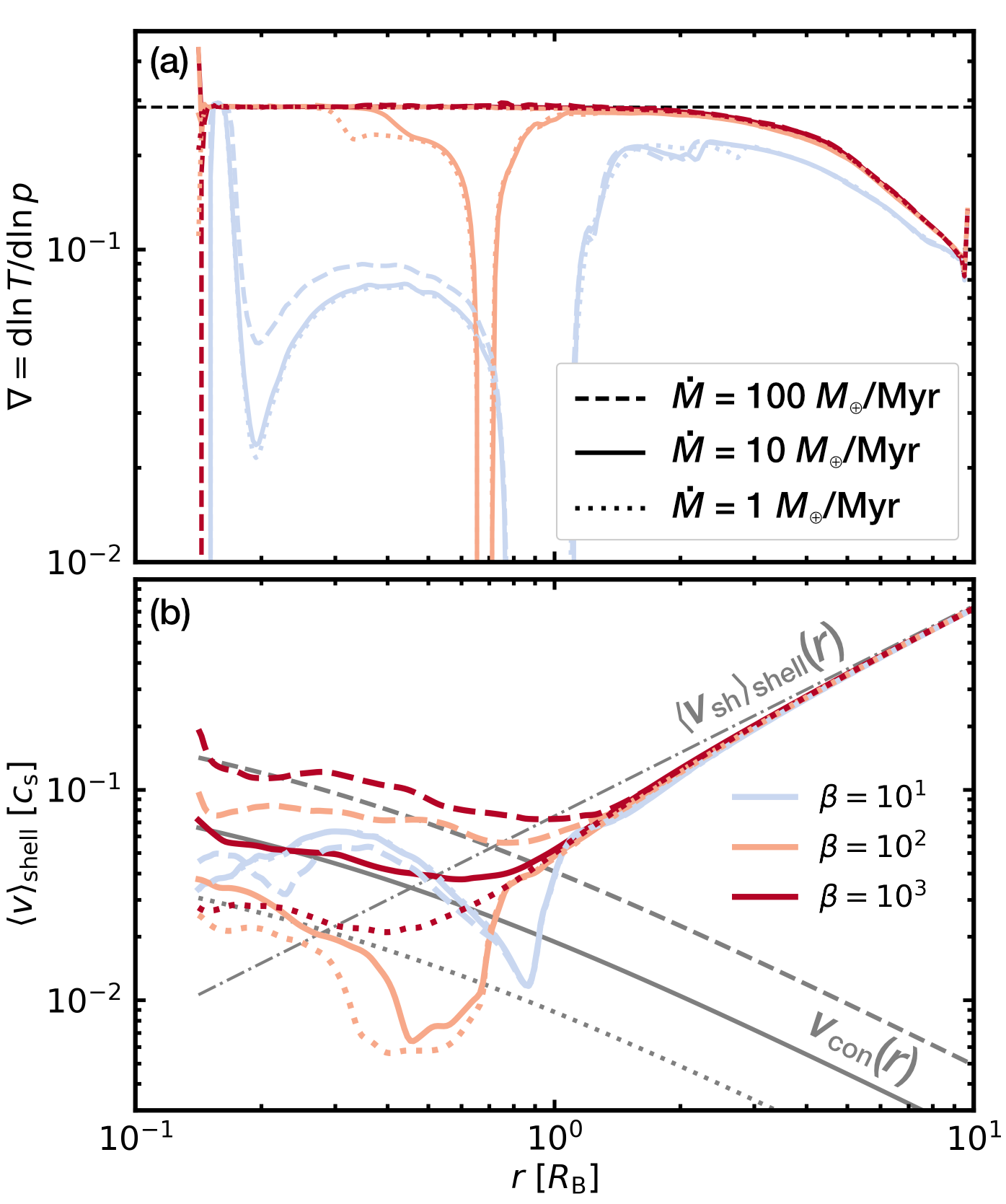}
    \caption{Shell-averaged temperature gradient and mean velocity for different $\beta$ values and accretion luminosities. All panels are the \editsec{states} at the end of the calculation, $t=50$. The gray dashed, solid, and dotted curves in panel b are given by \Equref{eq:vcon} with an assumption of fully convective envelopes ($l_{\rm m}=R_{\rm B}$). Thus, these gray lines should be compared with the red lines. The gray dashed-dotted curve is given by \Equref{eq:vsh_shell}.}
    \label{fig:Lacc_dependence}
\end{figure}

\begin{figure}[tp]
    \centering
    \includegraphics[width=1\linewidth]{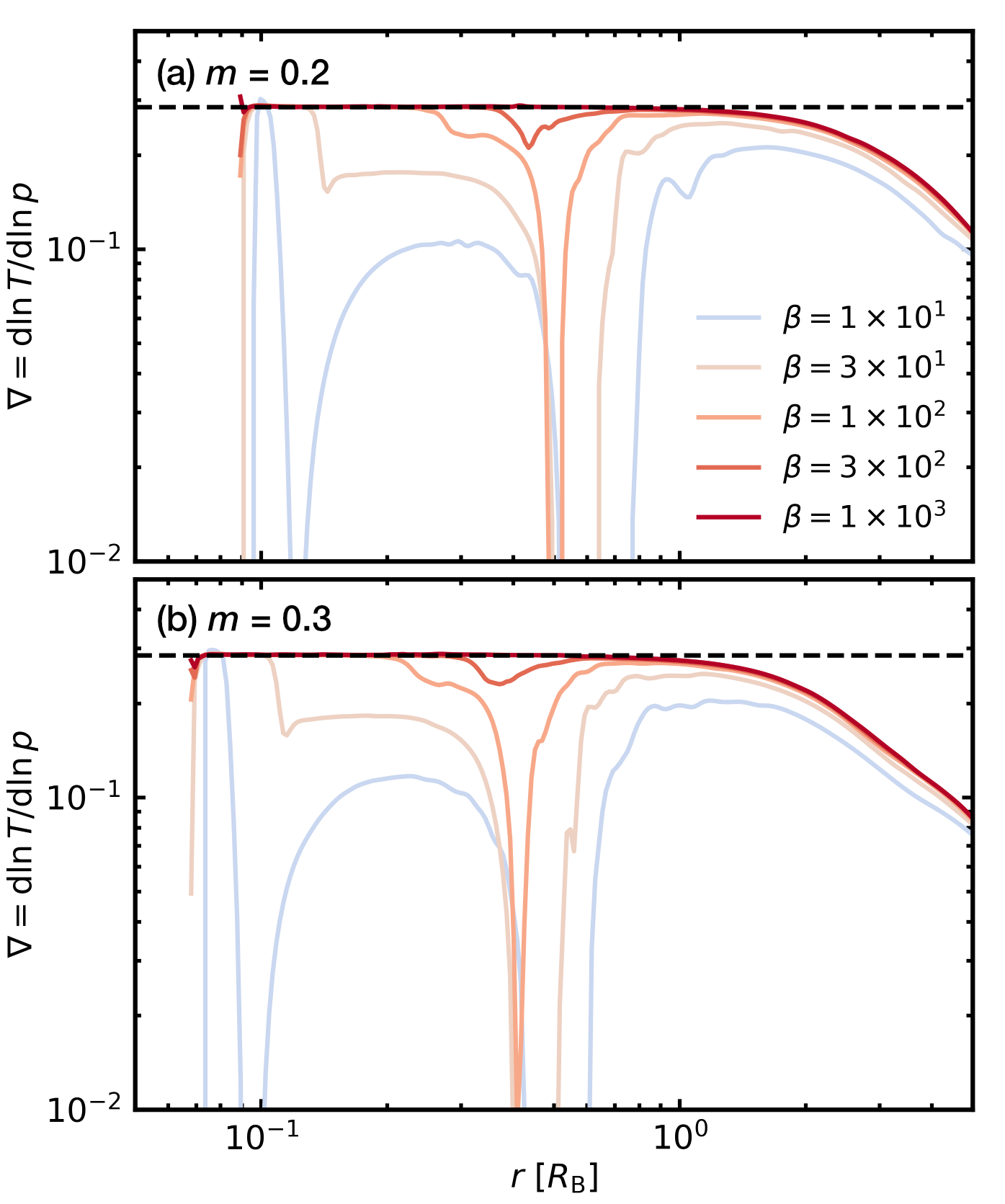}
    \caption{Shell-averaged temperature gradient for different planetary masses and $\beta$ values. We set $\dot{M}=10\,M_\oplus/$Myr. All panels are the snapshots at the end of the calculation, $t=50$. The horizontal dashed line is the adiabatic gradient, \Equref{eq:Schwarzschild criterion}.}
    \label{fig:Mp_dependence}
\end{figure}

Envelope dynamics \editsec{are} controlled by the balance of heating and cooling. \editsec{A} higher accretion luminosity steepens the temperature gradient, promoting convection. \editsec{In practice,} the accretion luminosity depends on planetary mass and pebble accretion rate (Eq. \ref{eq:Lacc}). \editsec{Therefore,} we vary the accretion rates by \editsec{two} orders of magnitude (\Tabref{tab:hydro simulations}). 
\editsec{From these numerical experiments, we observe that} the critical $\beta$ that separates the slow cooling regime from other regimes depends on the accretion luminosity. As shown in \Figref{fig:Lacc_dependence}a, full convection occurs when $\beta\geq10^3$ and $\dot{M}\leq10\,M_\oplus$/Myr, \editsec{but already at shorter cooling times ($\beta\geq10^2$) when $\dot{M}=100\,M_\oplus$/Myr.} \editsec{However, the} \edit{structure of the envelope does not change significantly with variations in the accretion luminosity. \editsec{We do note that}} the shell-averaged mean speed within the Bondi radius increases with the accretion luminosity, \editsec{as can be seen in \Figref{fig:Lacc_dependence}b}, as \editsec{expected from} Eq. \ref{eq:vcon}. \edit{\editsec{Finally, in the current work, we briefly explore the dependency on planetary mass.} We vary the planetary masses within a factor of \editsec{three}. \editsec{As expected, varying the planetary mass within a factor of three does not lead to substantial changes in the envelope structure (\Figref{fig:Mp_dependence}).}}

\begin{figure*}[t]
    \centering
    \includegraphics[width=1\linewidth]{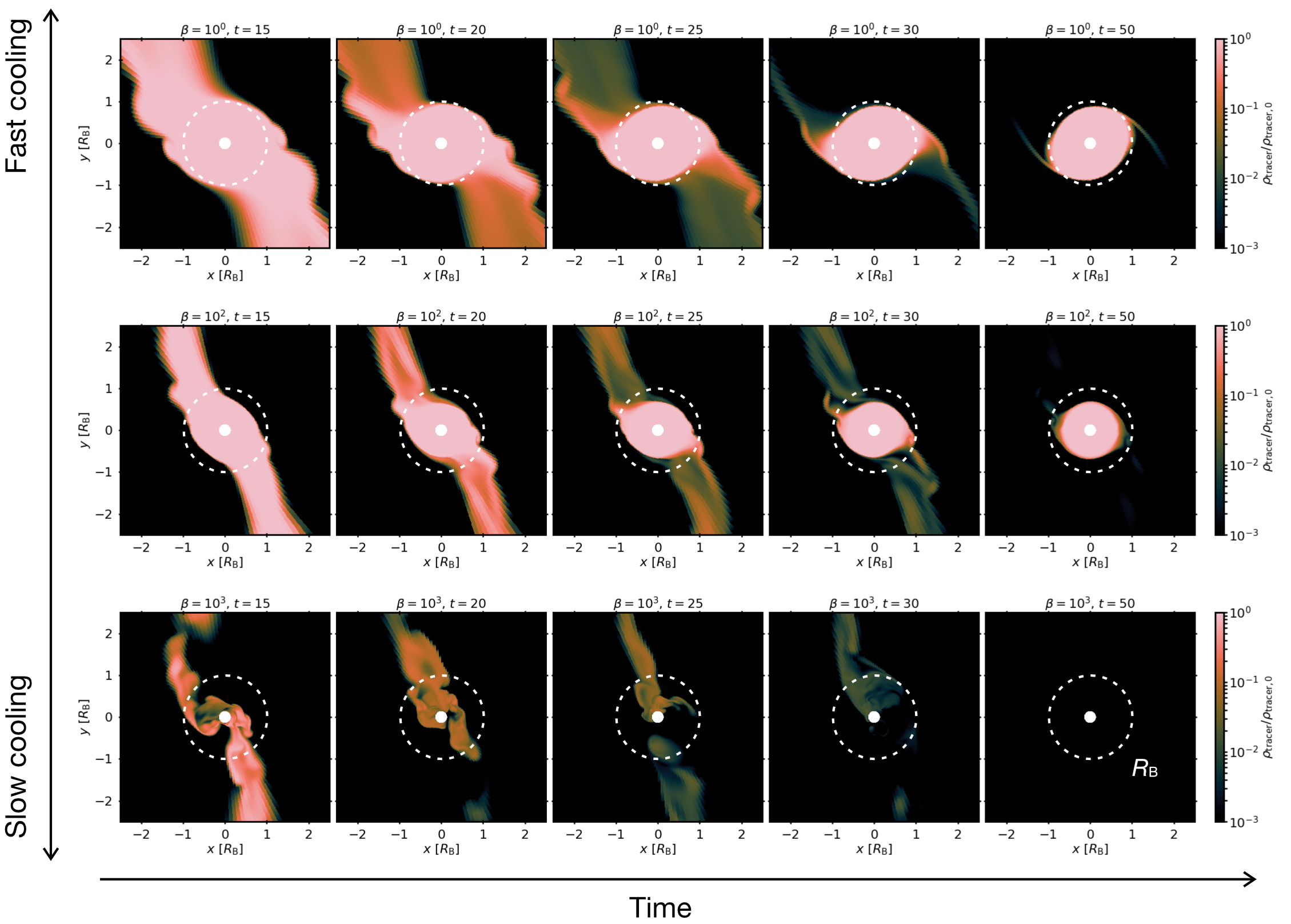}
    \caption{Tracer concentration at the midplane. We set $m=0.1$. Different rows correspond to different $\beta$-values ($\beta=10^{0},\,10^2,$ and $10^3$, from top to bottom). Different columns correspond to different time ($t=15,\,20,\,25,\,30,$ and 50 \editforth{in code units}, from left to right).}
    \label{fig:tracer_particle}
\end{figure*}

\begin{figure*}[t]
    \centering
    \includegraphics[width=1\linewidth]{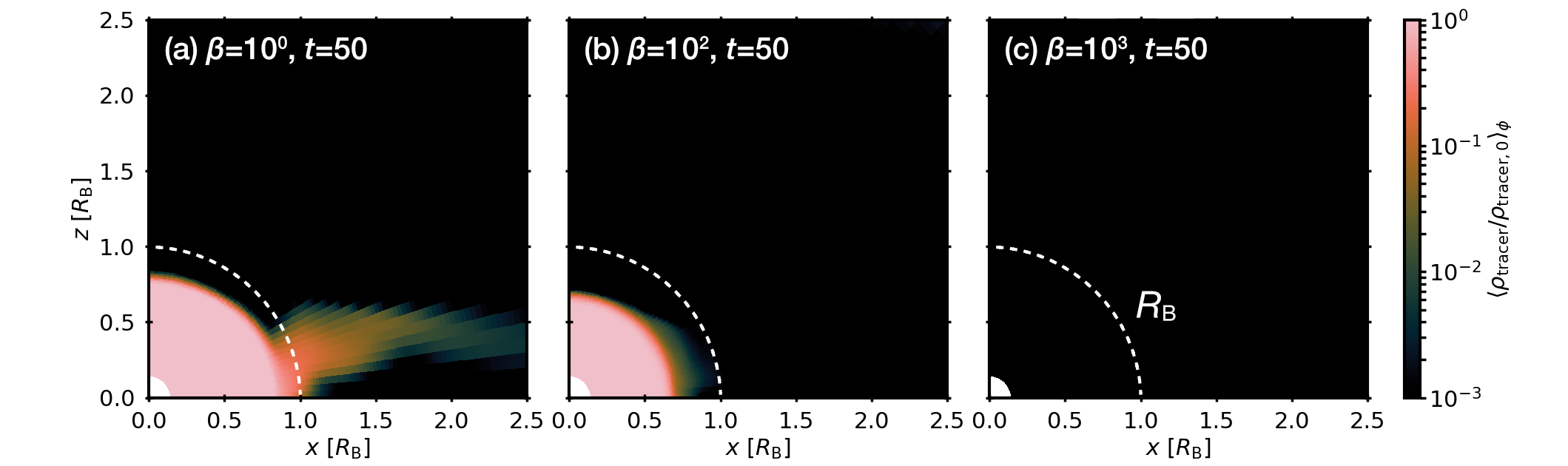}
    \caption{\editforth{Tracer concentration in the $x$-$z$ plane at the end of the calculation ($t=50$ in code units), after a hemispherical average along the azimuthal direction.}}
    \label{fig:tracer_particle_xz}
\end{figure*}

\begin{figure*}[t]
    \centering
    \includegraphics[width=1\linewidth]{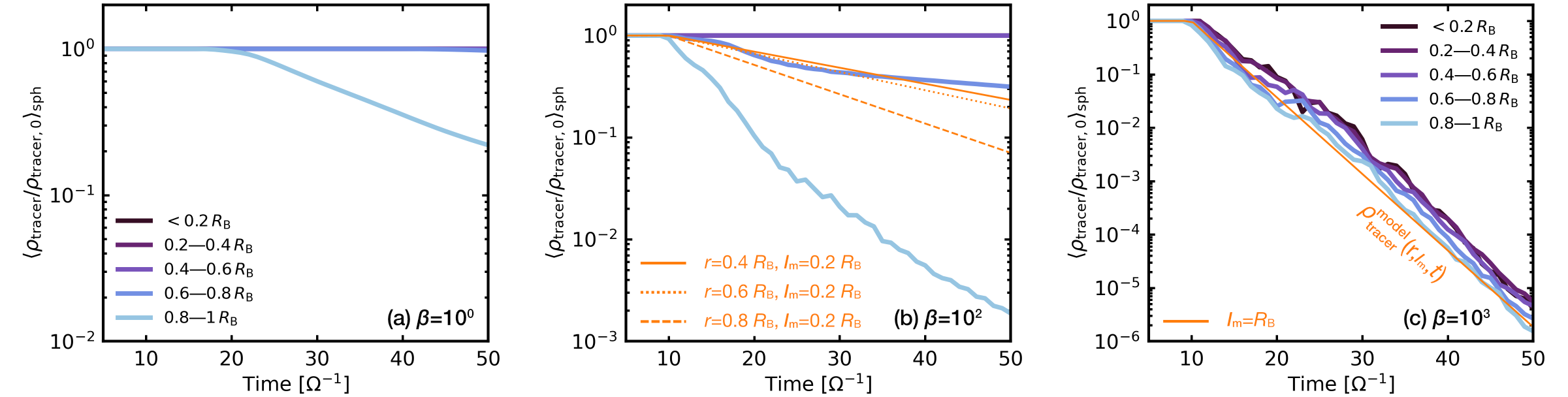}
    \caption{\editforth{Time evolution of the shell-averaged tracer concentration for $\beta=10^0$ (panel a), $10^2$ (panel b), and $10^3$ (panel c). Different color correspond to different shell radii. The orange lines are given by \Equref{eq:decay rate}. In panels b, we set $l_{\rm m}=0.2\,R_{\rm B}$ (partially convective) and plot the decay rate at $r=0.4\,R_{\rm B}$ (the solid line), $r=0.6\,R_{\rm B}$ (the dotted line), and $r=0.8\,R_{\rm B}$ (the dashed line), respectively. In panel c, we set $l_{\rm m}=R_{\rm B}$ (fully convective).}}
    \label{fig:tracer_particle_time}
\end{figure*}

\section{\edit{Tracing material transport}}\label{sec:Tracing material transport}

Envelope dynamics strongly influence the transport of small particles and vapors. To get an insight into material transport within the envelope, \edit{we use passive scalars as tracer particles in our hydrodynamical simulations \editsec{that obey} the following transport equation \citep{stone2020athena++}:
\begin{align}
    \frac{\partial \rho C_{\rm tracer}}{\partial t}+\nabla\cdot[\rho \bm{v}C_{\rm tracer}]=0.
\end{align}
We keep the passive scalar density \editsec{$C_{\rm tracer}$} equal to the gas density when $t<t_{\rm tracer,0}$. After $t\geq t_{\rm tracer,0}$\editsec{,} the passive scalar density varies in time with the gas motion. We set $t_{\rm tracer,0}=5\,\Omega^{-1}$. \editsec{A} reflecting inner boundary condition is applied to the passive scalars. For the outer boundary condition, we set the passive scalar density equal to 0 at $r>0.8\,r_{\rm out}$, \editsec{where $r_{\rm out}$ is the outer edge of the simulation domain (Sect. \ref{sec:Numerical method})}.}

\editsec{Figure \ref{fig:tracer_particle} shows the tracer concentration in the $x$-$y$ midplane for different cooling times, while \Figref{fig:tracer_particle_xz} illustrates the distribution in the $x$-$z$ perpendicular plane}. Tracers are confined within the radiative layer, which is isolated from the outer recycling layer by buoyancy. Outside the radiative layer, tracers are efficiently transported by recycling, shear, and horseshoe flows \editforth{(the top rows in Fig.~\ref{fig:tracer_particle} and \Figref{fig:tracer_particle_xz}a)}. Even if a convective layer lies beneath a radiative one, tracers remain trapped in the radiative layer \editforth{(the middle rows in Fig.~\ref{fig:tracer_particle} and \Figref{fig:tracer_particle_xz}b)}. However, in a fully convective envelope, tracers are expelled within a few tens of orbits \editforth{(the bottom rows in Fig.~\ref{fig:tracer_particle} and \Figref{fig:tracer_particle_xz}c)}.

\editsec{We explore further the time evolution of the tracer concentration as function of radius in the envelope, for different cooling times. \editforth{Figure \ref{fig:tracer_particle_time}} shows the tracer density in different radial zones as function of orbital time. \editforth{For the short $\beta$-cooling times ($\beta\lesssim1$), the tracer concentration remains nearly constant within the radiative layer, $r\lesssim0.6\text{--}0.8\,R_{\rm B}$}, decayling only in the outer recycling zone (\editforth{\Figref{fig:tracer_particle_time}a}).} In the intermediate cooling regime, $1\lesssim\beta\lesssim300$, in which the \edit{inner} envelope can be partially convective, the tracers stagnate in the innermost part of the envelope, $r\lesssim0.4\text{--}0.6\,R_{\rm B}$ (\editforth{\Figref{fig:tracer_particle_time}b}). In the slow cooling regime, $\beta\gtrsim1000$, the tracer concentration drops rapidly throughout the fully convective envelope (\editforth{\Figref{fig:tracer_particle_time}c}).

\edit{The evolution of the tracer concentration} \editsec{at a radius $r$ in the envelope} is well-described by a simple exponential decay:
\begin{align}
    \rho^{\rm model}_{\rm tracer}(r,l_{\rm m},\tilde{L}_{\rm acc},t)=\rho_{\rm tracer,0}\times\exp\Bigg(-\frac{t-t_{\rm tracer,0}}{\tau(r,l_{\rm m},\tilde{L}_{\rm acc})}\Bigg),\label{eq:decay rate}
\end{align}
where $\tau$ corresponds to the characteristic time scale to transport the tracers in the convective or radiative layer. The above equation is plotted in \editforth{Figs. \ref{fig:tracer_particle_time}b and c} with the orange lines, reproducing the numerical results. Here we define
\begin{empheq}
    [left={\displaystyle\tau(r,l_{\rm m},\tilde{L}_{\rm acc})\equiv\empheqlbrace}]{alignat=2}
    &\frac{2l_{\rm m}}{v_{\rm con}(r_{\rm in},l_{\rm m},\tilde{L}_{\rm acc})}\quad(\text{when}\,l_{\rm m}=R_{\rm B}),\nonumber\\
    &\frac{R_{\rm B}-r}{v_{\rm rad}}\quad(\text{when}\,0<l_{\rm m}<r< R_{\rm B}),\nonumber\\
    &\infty\quad(\text{when}\,r\leq l_{\rm m}<R_{\rm B}\,\text{or }l_{\rm m}=0),\label{eq:tau}
\end{empheq}
where $v_{\rm rad}$ is the mean flow speed in the radiative layer. When $m=0.1$, \Equref{eq:tau} yields $\tau\approx3\,\Omega^{-1}$ for a fully convective envelope\editsec{,} \edit{where the mixing length is comparable to the the Bondi radius} $(l_{\rm m}=R_{\rm B})$\editsec{, using \Equref{eq:vcon} for $v_{\rm con}$}. \edit{For a partially convective envelope, we find that \editsec{at depth of $r=0.4\,R_{\rm B}$} $\tau\approx30\,\Omega^{-1}$, taking $l_{\rm m}=0.2\,R_{\rm B}$.} These estimates rely on the following assumptions: 
(1) In a fully convective envelope, tracers are circulated over a distance of $2l_{\rm m}$ (circulation in a convective cell), regardless of their initial position.
(2) When the envelope has inner convective and outer radiative layers, tracers are transported outward in the radiative layer at a speed of $v_{\rm rad}=0.1\,v_{\rm con}(r,l_{\rm m},\tilde{L}_{\rm acc})$, inferred from \Figref{fig:shell_avg_vel_b1e2_1e3}.
(3) In a \edit{near-isothermal} envelope, the tracers are \edit{effectively} trapped. We note that even in a fully radiative envelope, radial velocities are small but finite ($10^{-5}\,c_{\rm s}$; \Figref{fig:shell_avg_vel_vr_vphi}), which implies a Bondi-crossing timescale \editsec{exceeding} $10^4$ orbits for a planet with $m=0.1$.



\begin{figure*}
    \centering
    \includegraphics[width=1\linewidth]{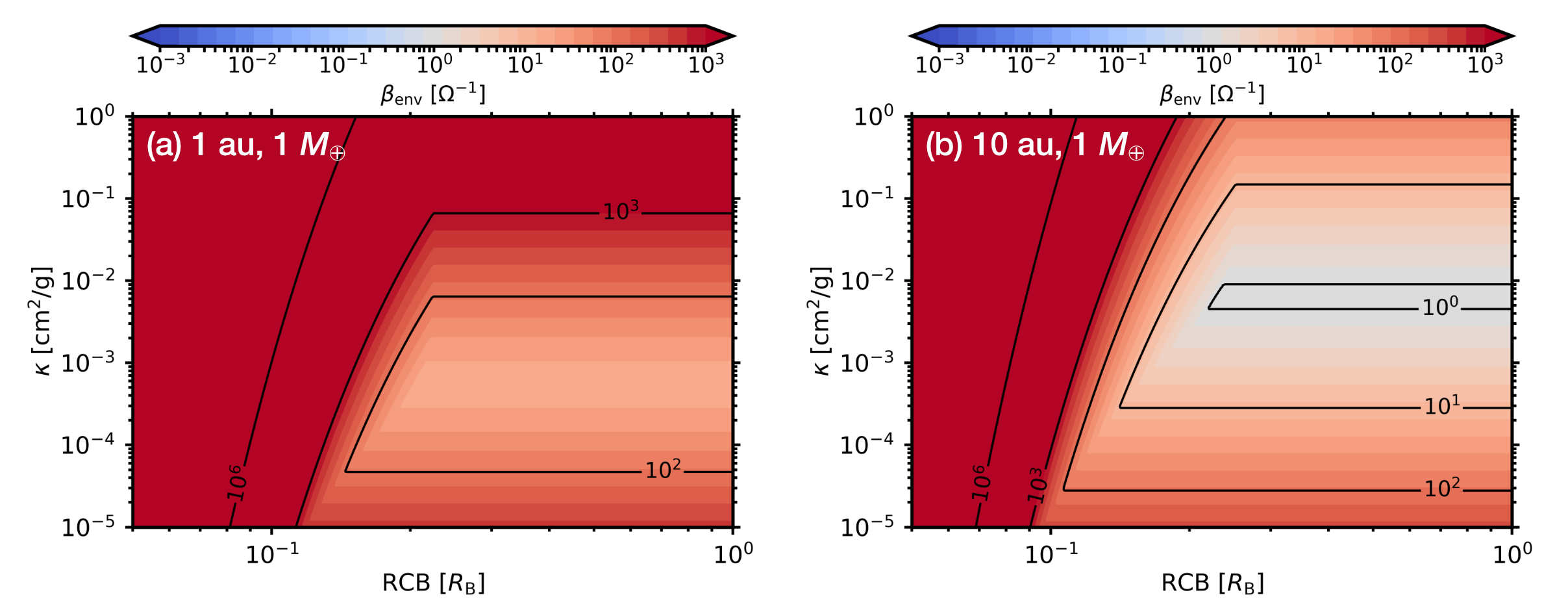}
    \caption{Analytic estimate of the envelope's cooling time as a function of the opacity and the RCB location. Different panels correspond to different orbital location of the Earth-mass planet.}
    \label{fig:beta_env}
\end{figure*}

\section{Discussions} \label{sec:Discussions}

\subsection{Model limitation: $\beta$-cooling assumption}
The $\beta$-cooling model employed in this study is a simplified radiative cooling model 
\citep[e.g.,][]{Gammie:2001}. \editsec{Assigning a fixed $\beta$ value throughout the envelope is an assumption that does not take into account that the primary source of opacity are small dust grains present in the envelope.} 
\edit{In Appendix \ref{sec:Photospheric radius}, we show that 
\editsec{the applicability of the $\beta$-cooling model} is valid when the opacity is low $\kappa\lesssim10^{-5}\text{--}10^{-4}\,\text{cm}^2\text{/g}$ or at distances outside of $\gtrsim$10 au.} \edit{However}, there is significant uncertainty in the opacity of planetary envelopes. The primary source of opacity is dust, and a commonly adopted value is $\kappa\sim1\,\text{cm}^2\text{/g}$, \edit{corresponds} to that of the interstellar medium (ISM). \editsec{However}, processes such as dust growth and settling can substantially reduce dust opacity, potentially resulting in values much lower than the ISM standard \citep{ormel2014atmospheric,mordasini2014grain}. Notably, in the region within $\lesssim0.1\,R_{\rm B}$, the opacity can drop to values as low as $\kappa\lesssim10^{-4}$ cm$^2$/g \edit{because of dust coagulation}. 


\subsection{Analytic estimate of envelope's cooling time}
\edit{The orbital location of the planet \editsec{determines the effective cooling rate of the envelope.} In Appendix \ref{sec:Cooling time estimation}}, we provide an analytical estimate of the envelope's cooling time, \edit{$\beta_{\rm env}$,} 
by combining the cooling time of the background disk gas---which constrains the cooling time at the outer edge of the envelope---with the depth-dependent cooling time derived from \editsec{a} two-layer envelope model. 
\edit{The cooling time of the background disk gas, $\beta_{\rm disk}$, takes into account \editsec{radiative cooling from the disk surface} and in-plane cooling \citep[radial thermal diffusion through the disk midplane;][]{ziampras2023modelling}. The cooling time derived from the two-layer model, $\beta_{\rm RCB}$, is the time it takes to cool the envelope until the radiative-convective boundary (RCB) reaches a given pressure depth \citep{piso2014minimum}.} 
\editsec{F}igure \ref{fig:beta_env} shows that, in 
the deep envelope ($r\lesssim0.1\,R_{\rm B}$) \editsec{that is not directly probed by our simulations}, the cooling time exceeds $10^3$ orbits, \editsec{with only a weak dependency on the dust opacity.} \edit{In contrast, in 
the outer envelope ($\gtrsim0.1\,R_{\rm B}$), \editsec{cooing times can be around 100 orbits for Earth-like planets in the inner disk (1 au, \Figref{eq:beta env}a), but decreasing with orbital radius to 1-100 orbits outside of 10 au (\Figref{eq:beta env}b).} \editsec{This indicates that}, envelopes will generally be in the fast or intermediate cooling regime in the outer disk, while this breaks down toward the inner disk where cooling timescales are long. There, planets approaching the Earth to super-Earth regime, would acquire fully convective envelopes. This would hamper further growth, as such low-mass envelopes have long growth timescale \citep{Lambrechts:2017} and even pebble accretion rates may be reduced (see Sect. \ref{sec:Impact of envelope dynamics on planetary growth}).}


Our simulations primarily cover the domain outside $0.1\,R_{\rm B}$, \editsec{where} the cooling time is \editsec{mainly} set by $\beta_{\rm disk}$. If the opacity $\kappa$ is \editsec{approximately constant in this region,} then $\beta_{\rm env}$ is also constant throughout the domain (\Figref{fig:beta_env}), \editsec{in line with a the uniform $\beta$ value in our simulation domain.} However, the opacity within the envelope can vary by several orders of magnitude due to processes such as dust growth and settling \citep{ormel2014atmospheric,mordasini2014grain}. Moreover, gas dynamics may cause the opacity to vary not only radially\editsec{,} but also vertically and azimuthally, resulting in 3D \editsec{variations} within the envelope \citep{krapp20223d}. The co-evolution of dust and the envelope is further discussed in Sect. \ref{sec:Sustainability of dynamical states in envelopes}.

\subsection{Comparison to previous studies}
Numerous simulations of \edit{the envelopes of} embedded planet have been conducted, both with and without accretion luminosity. In this section, we first compare our results with previous studies that neglect accretion luminosity (Sect. \ref{sec:Without luminosity}), followed by those that include it (Sect. \ref{sec:With luminosity}).

\subsubsection{Without accretion luminosity}\label{sec:Without luminosity}
Although neglecting the accretion luminosity is an unrealistic assumption and requires careful treatment in numerical simulations \citep{zhu2021global}, it remains a convenient simplification and has been widely adopted in previous studies.

In the isothermal limit, where buoyancy forces are absent, polar inflow can penetrate deep into the envelope resulting in a structure dominated entirely by the recycling layer \citep{Ormel:2015b,Fung:2015,Kuwahara:2019}. As shown in \editforth{\Figref{fig:envelope_layer}, smaller values of $\beta$ lead to radially wider recycling layers}, consistent with these earlier findings. In our simulations, the lower bound of the cooling time is $\beta = 0.01$; we expect that even smaller $\beta$ would reproduce the isothermal limit.

In reality, envelopes have a finite cooling time. Using $\beta$-cooling model with $\beta\leq1$, \cite{Kurokawa:2018} conducted 3D hydrodynamic simulations and found that the radiative layer is isolated from the recycling layer due to a buoyancy force, and that its extent increases with $\beta$. This is consistent with our results \editforth{(Fig.~\ref{fig:envelope_layer})}.

In the absence of convection, envelope recycling is more efficient in the outer \edit{envelope} regions, a trend consistently observed in both isothermal and non-isothermal envelopes \citep{Ormel:2015b,Cimerman:2017,Bethune:2019,moldenhauer2021steady,bailey2024growing}. \cite{Cimerman:2017} performed 3D radiation hydrodynamic simulations using the flux-limited diffusion (FLD) approximation for Earth-like planets, assuming \edit{a} low opacity ($\kappa=10^{-4}\text{--}10^{-2}\,\text{cm}^2\text{/g}$). They showed that the outer envelope ($r\gtrsim0.4\,R_{\rm B}$) recycles within approximately 1 to 10 orbits, while the inner regions recycle over approximately 100 orbits. A subsequent study confirmed that the entire envelope recycles over finite timescales, even if the inner radiative layer is isolated from recycling \citep{moldenhauer2022recycling}. \cite{moldenhauer2022recycling} reported that the bulk of the envelope recycles on a timescale of approximately $10^2$ orbits, except for regions extremely close to the planetary surface, where timescales can be as long as $10^4$ orbits. Although the physical mechanism remains unclear, \edit{the authors} noted that the envelope becomes turbulent, which enables even deeply embedded tracers to eventually escape.  
\edit{In contrast,} our results suggest lower recycling efficiency in the inner envelope. \editforth{As discussed in Section~\ref{sec:Tracing material transport}, the recycling timescales in the radiative layer can exceed $10^4$ orbits.} No prominent turbulence is observed in the radiative layer. Since hydrostatic equilibrium is established, radial velocities remain very small (\Figref{fig:shell_avg_vel_vr_vphi}). A recent simulation solving the radiative transfer equations has also confirmed that the inner envelope ($r < 0.4\,R_{\rm B}$) remains insulated from recycling \citep{bailey2024growing}, consistent with our findings.

\edit{The} efficiency of recycling in the radiative interior continues to be debated, especially in light of the diversity of numerical treatments across different models. In regions near the inner boundary, where density and pressure gradients become steep, either high numerical resolution or large gravitational smoothing lengths are required to maintain hydrostatic balance \citep{zhu2021global}. Insufficient resolution in these regions leads to artificial velocity fluctuations and numerical heating, which affects the entire envelope \citep{Ormel:2015b,zhu2021global}. Furthermore, \cite{zhu2021global} noted that an explicit luminosity input is required to accurately model the envelopes. Therefore, the accretion luminosity should not be neglected.

\subsubsection{With accretion luminosity}\label{sec:With luminosity}
The envelope of an embedded planet is inherently heated by accretion of solids and compressional heating of the gas, resulting in intrinsic luminosity. Several sophisticated radiation-hydrodynamic simulations of envelopes have been carried out so far. \cite{Lambrechts:2017} performed 3D radiation hydrodynamic simulations using the FLD approximation and found that the envelope exhibits a three-layer structure: the convective, radiative, and recycling layers from inside out. Based on their adopted parameters---$\Sigma_0=400\,\text{g/cm}^2$, $T_0=40$ K, $M_{\rm p}=5\,M_\oplus$, $a_{\rm p}=5.2$ au, and $\kappa=0.01\,\text{cm}^2\text{/g}$---the envelope cooling time exceeds $10^3$ orbits only in the innermost region ($r<0.3\,R_{\rm B}$); beyond $0.3\,R_{\rm B}$, $\beta_{\rm env}$ is approximately 20, placing it within our intermediate cooling regime. In this regime, we also find a three-layer envelope structure, which is consistent with their results.

\cite{Popovas:2018b} carried out 3D radiation hydrodynamic simulations, finding that the envelope is fully convective, with convective velocities ranging from approximately 0.1 to 1 km/s (corresponding to $0.1\text{--}1\,c_{\rm s}$). These velocities are higher than ours (typically smaller than $0.1\,c_{\rm s}$; \Figref{fig:shell_avg_vel_b1e2_1e3}), due to their assumption of a low disk gas density. Since $v_{\rm con}\propto \rho^{-1/3}$ (Eq. \ref{eq:vcon}), lower density naturally leads to higher convective velocities. Using their fiducial parameters---$M_{\rm p}=1\,M_\oplus,\,L_{\rm acc}=2\,M_\oplus\text{/Myr},\,\rho_0=10^{-10}\,\text{g/cm}^2$, and $T=300\,\text{K}$--in our analytic model yields velocities of $0.1\text{--}1\,c_{\rm s}$, which is in excellent agreement with their findings.

\cite{zhu2021global} directly solved time-dependent radiative transfer equations using the radiation module of the Athena++ code. They adopted a detailed opacity table, in which the opacity varies with depth, but assumed $\kappa \sim 1\,\text{cm}^2\text{/g}$ and $0.1\,\text{cm}^2\text{/g}$ in the outer envelope.
Using their parameters, $\Sigma_0=488\,\text{g/cm}^2$, $T_0=39.4$ K, $M_{\rm p}=10\,M_\oplus$, and $a_{\rm p}=5$ au, we estimate $\beta_{\rm env}\approx2000$ for $\kappa=1\,\text{cm}^2\text{/g}$ and $\beta_{\rm env}\approx200$ for $\kappa=0.1\,\text{cm}^2\text{/g}$ in the region beyond $0.5\,R_{\rm B}$ \edit{(Eq. \ref{eq:beta env})}. According to our results, $\beta_{\rm env}=2000$ corresponds to a fully convective envelope, while $\beta_{\rm env}=200$ results in a radiative layer forming above a convective interior. These expectations are consistent with their findings: the envelope is fully convective and rapidly recycles tracers for $\kappa=1\,\text{cm}^2\text{/g}$, while a radiative layer develops and inhibits recycling in the inner region ($r\lesssim0.2\,R_{\rm B}$) for $\kappa=0.1\,\text{cm}^2\text{/g}$.


Finally, we note that accretion heating can alter the 3D gas flow structure. \cite{Chrenko:2019} performed 3D radiation hydrodynamic simulations using the FLD approximation. They found that disk gas enters the Hill sphere through the midplane and exits through the polar regions. Their setup---$\Sigma_0=484\,\text{g/cm}^2$, $T_0=81$ K, $M_{\rm p}=3\,M_\oplus$, and $a_{\rm p}=5.2$ au, and $\kappa=1.1\,\text{cm}^2\text{/g}$---corresponds to $\beta_{\rm env}\approx50$ \edit{(Eq. \ref{eq:beta env})}. They showed a preferential vertical (+$z$) outflow at the Bondi scale, although their resolution (8 grids/$R_{\rm H}$) did not resolve the envelope interior in detail. We observe a similar trend in our results (see \editforth{Figs. \ref{fig:vr}d} and \ref{fig:azimuth_and_time_average}c): for larger $\beta$ values, the radially outward velocity becomes more prominent above the midplane.

\subsection{Model limitation: pure Keplerian rotation}\label{sec:Model limitation: pure Keplerian rotation}
In this study, we neglect the pressure gradient of the background disk gas and assumed  pure Keplerian rotation. We believe that this simplification is expected to have a minor impact on our results \edit{\citep{moldenhauer2022recycling}.} In reality, the disk gas rotates slower than the Keplerian speed due to the radial pressure gradient (sub-Keplerian rotation), and the planet experiences a headwind from the disk gas. This headwind \editsec{could reduce} the size of the \edit{radiative layer} of the envelope. 

According to a fitting model that includes the effect of headwind on atmospheric radius, the boundary between the recycling and radiative layers is given by \cite{kuwahara2024analytic},
\begin{align}
    R_{\rm atm}^{\rm KK24}=R_{\rm atm,0}\Bigg(1-\frac{D_1}{C_1}\frac{\eta v_{\rm K}/c_{\rm s}}{R_{\rm B}/H}\Bigg).\label{eq:Ratm,fit}
\end{align}
Here \edit{$R_{\rm atm,0}=C_1\,R_{\rm B}$} is the atmospheric radius without the effect of the headwind, $C_1=0.84$ and $D_1=0.056$ are the fitting coefficients, $\eta$ is a dimensionless quantity characterizing the global pressure gradient of the disk gas, and $v_{\rm K}$ is the Keplerian speed. \editsec{T}ypically, the headwind velocity is $\eta v_{\rm K} \lesssim 0.1\,c_{\rm s}$ across a wide range of orbital radii. For a planet with a thermal mass of $m=0.1$, we have $R_{\rm B}/H=0.1$. \editsec{Under} these conditions, the reduction in atmospheric size due to the headwind is up to approximately 7\% (Eq. \ref{eq:Ratm,fit}).

\subsection{Volatile delivery via pebble accretion}

Envelope dynamics strongly regulate the delivery and retention of volatile species, such as water, during planet formation \citep[e.g.,][]{johansen2021pebble}. \edit{If these species sublimate form accreting pebbles at a radius $R_{\rm sub}$ in the envelope, and this layer is located in a convective \editsec{or} recycling layer, sublimated vapor would be redistributed and potentially lost to the disk via advection though the outer envelope \citep{wang2023atmospheric}.} 

\begin{figure}[t]
    \centering
    \includegraphics[width=1\linewidth]{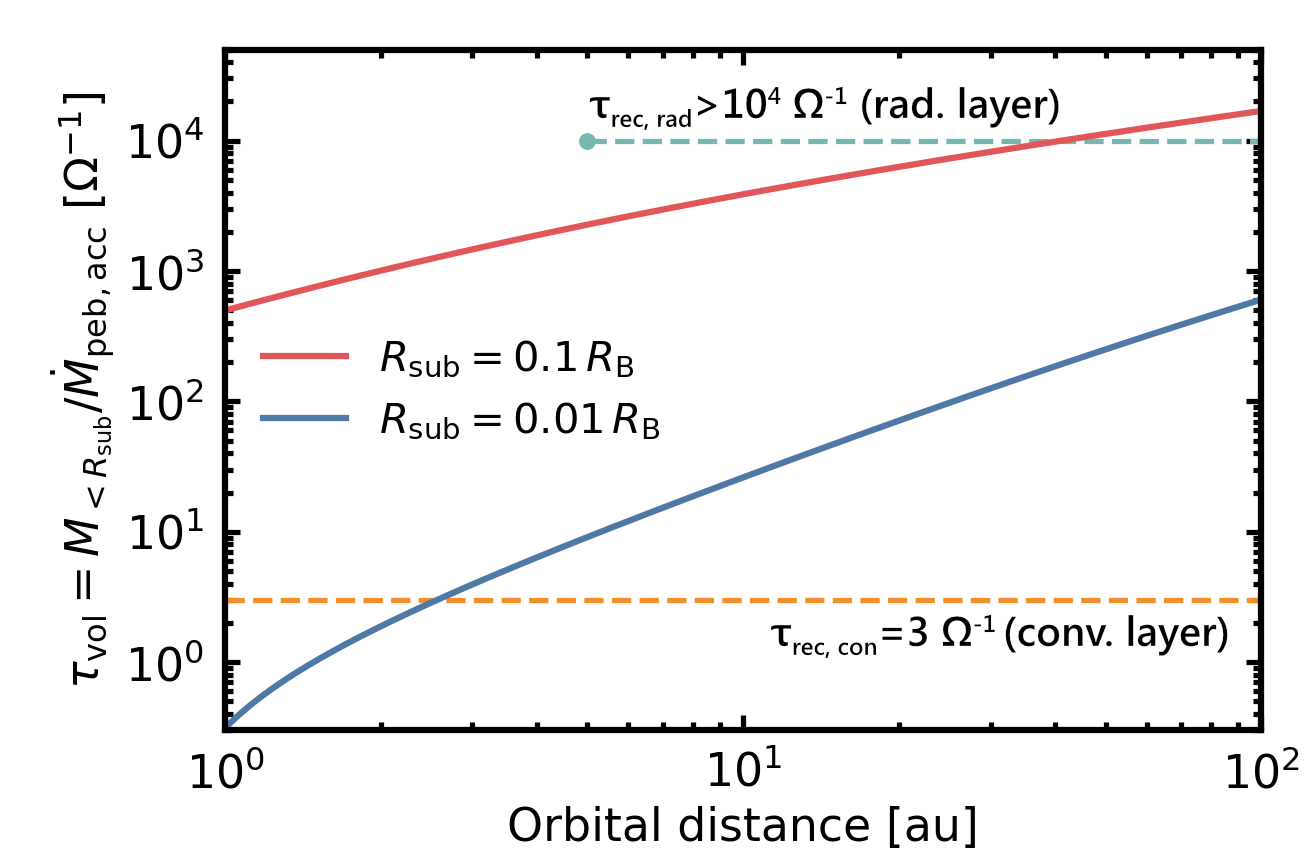}
    \caption{\edit{Volatile enrichment timescale as function of orbital distance. In the inner disk ($<10$ au), volatiles sublimating at \editthird{a} depth of approximately $0.1\,R_{\rm B}$ (red line) are efficiently lost through convective mixing ($\tau_{\rm rec,con}$, dashed orange line). In the colder outer disk, volatiles sublimate at greater depth in the envelope: at a depth of $0.01\,R_{\rm B}$ (blue curve) \editsec{volatiles} are typically deposited in \editsec{the} radiative zone \editsec{of the envelope where recycling timescales are long ($\tau_{\rm rec,rad}$, dashed lightblue line), preventing material exchange. \editsec{The lightblue dashed line is truncated at 5 au, since we expect that the envelopes are prone to become \editthird{fully} convective in the inner disk (\Figref{fig:beta_env}).}}}}
    \label{fig:tau_vol}
\end{figure}

\editsec{A}ssuming that volatiles are supplied via instantaneous sublimation of accreting icy pebbles, the timescale to enrich the envelope in volatiles can be expressed as:
\begin{align}
    \tau_{\rm vol}\equiv\frac{M_{<R_{\rm sub}}}{\dot{M}_{\rm peb,acc}},\label{eq:tau vol definition}
\end{align}
where $\dot{M}_{\rm peb,acc}$ is the pebble accretion rate and 
\begin{align}
    M_{<R_{\rm sub}}=\int_{R_{\rm p}}^{R_{\rm sub}}4\pi r^2\rho(r)\mathrm{d}r,\label{eq:MRsub}
\end{align}
is the mass between the core surface and the sublimation radius of \editsec{the considered species}. \edit{We numerically integrate \Equref{eq:MRsub} assuming $\rho(r)=\rho_{\rm ad}(r)$. We then assumed a passively irradiated disk model (Eq. \ref{eq:oka disk model}). \editsec{For the solid accretion rate we assume a standard accretion rate of} $\dot{M}_{\rm peb,acc}=10\,M_\oplus$/Myr in \Equref{eq:tau vol definition} \editsec{(the actual fraction may of course be smaller fraction of the total solid accretion rate for trace species).} The red line in \Figref{fig:tau_vol} shows the volatile enrichment timescale for $R_{\rm sub}=0.1\,R_{\rm B}$, which is appropriate for water \citep{wang2023atmospheric}. \editsec{For the case of water ice sublimation, we} note that the sublimation radius shrinks due to the cooling caused by ice sublimation \citep{wang2023atmospheric}, and \editsec{sublimation depths vary} for different volatile species. Thus, in \Figref{fig:tau_vol}, we also plot $\tau_{\rm vol}$ with $R_{\rm sub}=0.01\,R_{\rm B}$ as the blue line.} 

\begin{figure}[t]
    \centering
    \includegraphics[width=1\linewidth]{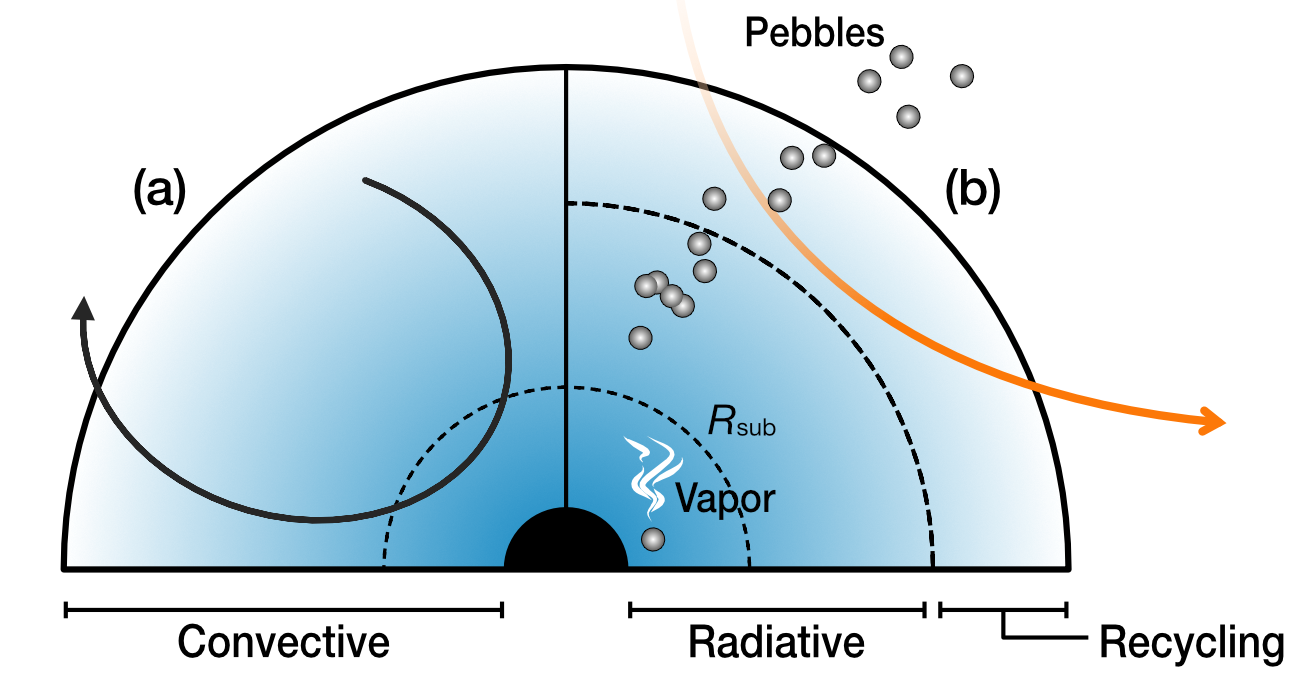}
    \caption{Schematic illustration of \editsec{(a) an envelope with a convective envelope (left side) and (b) an envelope with a shielding radiative layer and an outer recycling layer (right side).}}
    \label{fig:schematic}
\end{figure}

This timescale must be compared to the recycling timescale to assess whether volatiles are retained or lost. \editsec{\cite{wang2023atmospheric} showed that the fate of volatiles depends on the location of the sublimation front relative to the bound envelope: volatiles are trapped when sublimation occurs inside the bound region but are lost along the horseshoe flows if it occurs farther out. Although our simulations include 3D recycling flows absent in \edit{their} 2D models, the qualitative trends remain consistent.} In fully convective envelopes (e.g., inner disk regions), $\tau_{\rm rec,con}\sim3\,\Omega^{-1}$ \edit{(Sect. \ref{sec:Tracing material transport})}, which is much shorter than $\tau_{\rm vol}$. Consequently, sublimated volatiles are rapidly recycled back to the disk, making \edit{the planets forming in inner disk regions} volatile-poor \edit{(Figs. \ref{fig:tau_vol} and \ref{fig:schematic}a)}. In contrast, in envelopes with the radiative layer (e.g., outer disk regions)---especially at $r\lesssim0.4\text{--}0.6\,R_{\rm B}$---the recycling timescale exceeds $\tau_{\rm rec,rad}>10^4$ orbits. \edit{Because the sublimation radius shrinks by an order of magnitude in the outer disk regions \editsec{due to colder disk midplane temperatures} \citep{wang2023atmospheric}, we expect $\tau_{\rm vol}<\tau_{\rm rec,rad}$ (\Figref{fig:tau_vol}). \editsec{This} allows volatiles to accumulate in the deep interior \edit{(\Figref{fig:schematic}b)}. Therefore, planets forming in colder, outer disk regions ($\gtrsim10$ au) are more likely to become volatile-rich.}



\subsection{Impact of envelope dynamics on planetary growth}\label{sec:Impact of envelope dynamics on planetary growth}
Envelope dynamics \editsec{impact pebble accretion rates}. Recycling flows can suppress the accretion of small pebbles by deflecting them away from the planetary core \citep{Kuwahara:2020a,Kuwahara:2020b,okamura2021growth}. Convective motions can further disrupt pebble trajectories \citep{Popovas:2018a,Popovas:2018b}.

Although not modeled in this study, pebbles entering the envelope are expected to undergo thermal ablation \citep{Alibert:2017,brouwers2018cores}. When gas recycling is efficient, vaporized material is likely to escape the envelope, limiting core growth \citep{Alibert:2017}. However, when radiative layers are more stable and recycling is less efficient, metal-rich material may be retained, possibly contributing to envelope enrichment \citep{brouwers2018cores,brouwers2020planets,ormel2021planets}. Moreover, processes such as pebble erosion, fragmentation, and growth may significantly alter pebble transport and retention within the envelope \citep{ali2020limits,johansen2020transport}. Future studies should aim to couple hydrodynamic simulations with detailed pebble dynamics to better understand the co-evolution of pebble and envelope during planet formation.

\subsection{Sustainability of dynamical states in envelopes}\label{sec:Sustainability of dynamical states in envelopes}
The dynamical state of an envelope varies significantly depending on the cooling efficiency of the gas. If dust is the dominant source of opacity within the envelope, this leads to a fundamental question: \edit{can} the convective or radiative layer continue to be sustained?

Let us consider an initially fully convective envelope. In this case, the entire envelope is expected to be recycled in a short timescale. Since small dust particles, which contribute to large opacity, are tightly coupled to the gas, convection may reduce the abundance of small dust within the envelope, leading to a decrease in opacity. As a result, radiative cooling may become more effective, suppressing convection. Meanwhile the planet continues to accrete pebbles. Through processes such as \edit{pebble} fragmentation within the envelope, the abundance of small dust grains \edit{could be enhanced}, thereby increasing the opacity and re-initiating convection. The dust supply rate into the envelope depends on factors such as dust size and the 3D structure of the recycling layer. Ultimately, future multi-fluid (gas and dust) 3D simulations coupled with dust growth \edit{and fragmentation} are needed to understand the complete evolution and structure of the envelope in detail.


\section{Conclusions} \label{sec:Conclusions}
We conducted a suite of three-dimensional hydrodynamical simulations to characterize the structure and dynamics of planetary envelopes under a wide range of radiative cooling and accretion heating conditions. By varying the dimensionless cooling timescale $\beta$, we identified three distinct envelope regimes: fast cooling ($\beta\lesssim1$), intermediate cooling ($1\lesssim\beta\lesssim300$), and slow cooling ($\beta\gtrsim1000$). Each regime exhibits \edit{a characteristic envelope structure.} In the fast cooling regime, the envelope is nearly isothermal, featuring an inner radiative layer largely shielded from disk gas recycling. In the intermediate cooling regime, the envelope develops a three-layer structure comprising an inner convective layer, a middle radiative layer, and an outer recycling layer. In the slow cooling regime, the entire envelope is fully convective, which allows efficient recycling throughout \editsec{the envelope}.

We developed analytic models for convective and radiative transport timescales, which successfully reproduce the tracer particle evolution in each cooling regime. \edit{These models demonstrate that fully convective envelopes exchange material within approximately 3 orbits, while envelopes with radiative layers exchange material on \editsec{longer} timescales exceeding $10^4$ orbits.} 

\edit{By comparing the volatile enrichment timescale $\tau_{\rm vol}$ with the material exchange timescale $\tau_{\rm rec}$,} we found that envelopes formed in the inner disk ($\lesssim$1 au) are prone to become fully convective. \editsec{In such envelopes, the volatile enrichment timescale $\tau_{\rm vol}$ exceeds  the material exchange timescale $\tau_{\rm rec}$, implying these planets will be volatile poor.} 
Conversely, in the outer disk ($\gtrsim$10 au), radiative layers are more likely to develop \edit{at $r\lesssim0.4\text{--}0.6\,R_{\rm B}$, for which $\tau_{\rm vol}<\tau_{\rm rec}$}, enabling the accumulation of \edit{sublimated} volatiles in the envelope’s interior. This spatial variation implies a potential dichotomy in planetary compositions depending on where in the disk a planet forms.

\edit{To conclude, our work highlights the need to consider both envelope gas dynamics and thermodynamics} when assessing planetary growth and composition. \edit{We hope this motivates the development of} future models that incorporate dust evolution, radiative transfer, and multi-fluid dynamics to better understand the co-evolution of gas, solids, and volatiles \edit{during planet formation}.


\section*{Data Availability}

\begin{acknowledgements}
\editforth{We would like to thank an anonymous referee for the positive and encouraging report, that has improved the quality of this manuscript.} 
We thank \editsec{the} Athena++ developers. The Tycho supercomputer hosted at the SCIENCE HPC center at the University of Copenhagen was used for supporting this work. \editsec{M.L. acknowledge the ERC starting grant 101041466-EXODOSS.} Helpful discussions with Ziyan Xu.
\end{acknowledgements}
%


\bibliography{Ref_aanda}

\clearpage
\begin{appendix}
\def\thesection{A}
\setcounter{equation}{0}
\def\theequation{A.\arabic{equation}}
\setcounter{figure}{0}
\def\thefigure{A.\arabic{figure}}

\begin{figure}[htbp]
    \centering
    \includegraphics[width=1\linewidth]{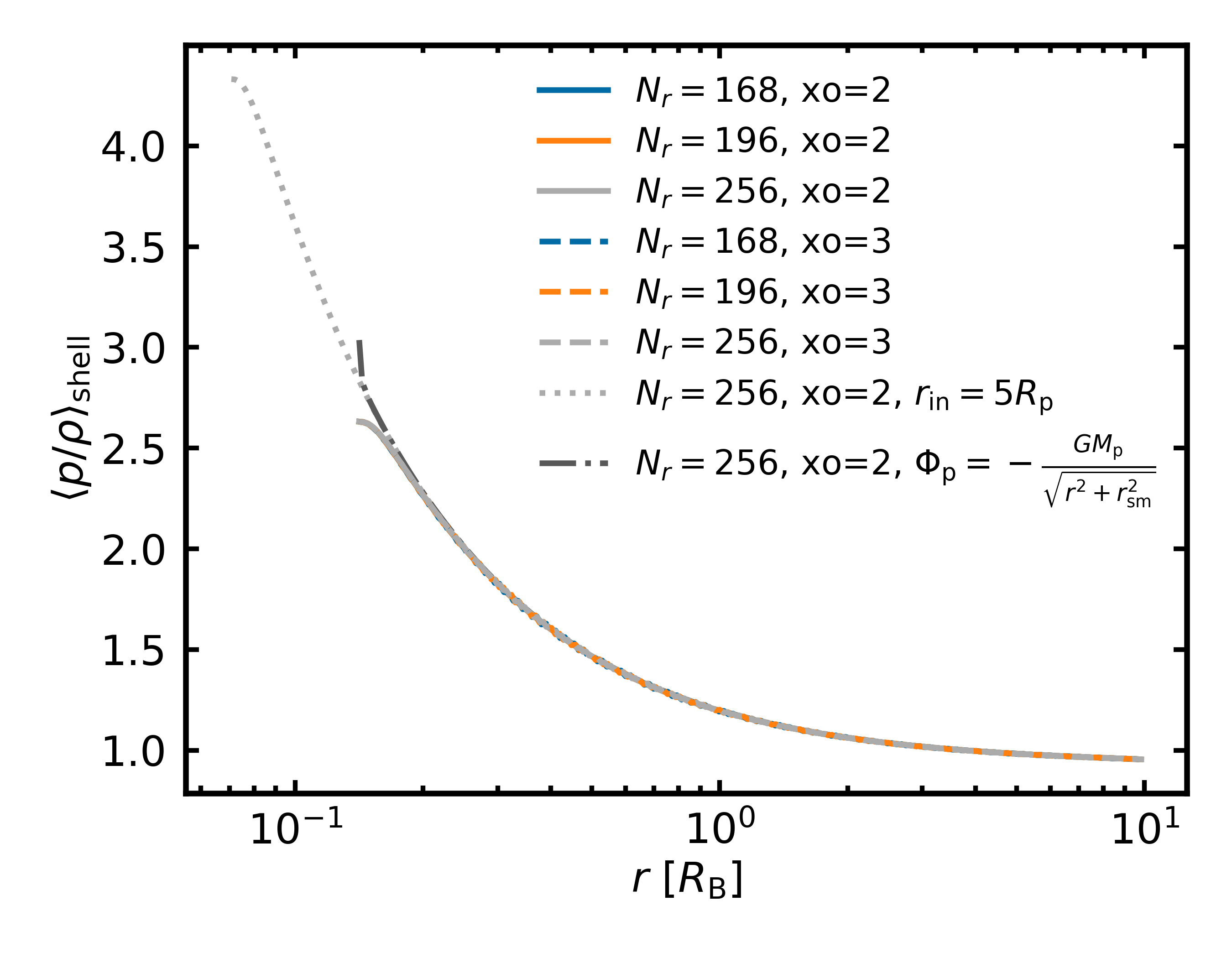}
    \caption{Sell-averaged temperature for different radial resolution, inner boundary, spatial reconstruction method, and potential formula. We set $m=0.1$ and $\beta=10^3$.}
    \label{fig:appendix sph_avg_temp}
\end{figure}
\begin{figure}[htbp]
    \centering
    \includegraphics[width=1\linewidth]{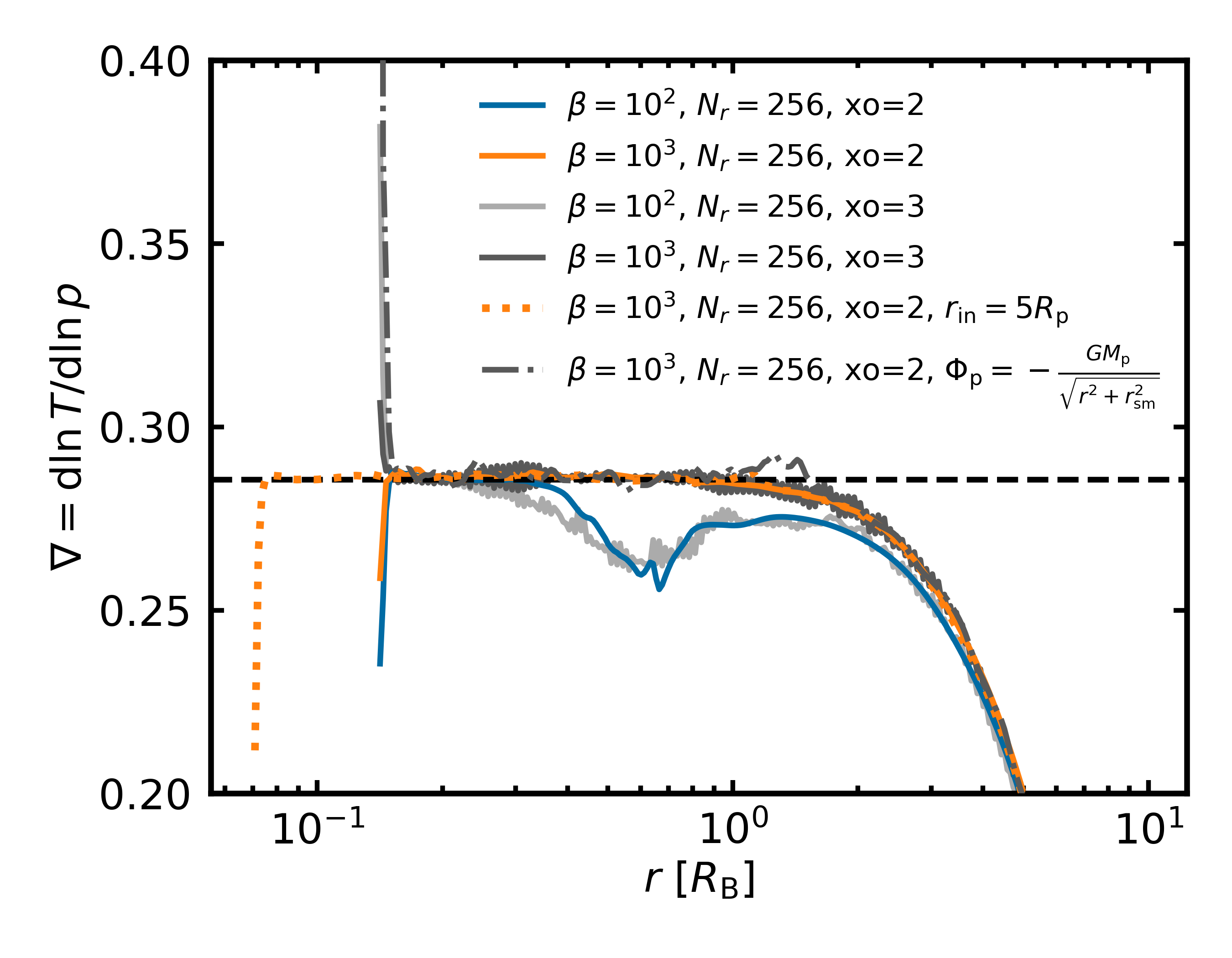}
    \caption{Shell-averaged temperature gradient for different $\beta$, inner boundary, spatial reconstruction method, and potential formula. We set $m=0.1$ and $N_r=256$.}
    \label{fig:appendix grad}
\end{figure}

\begin{figure}[htbp]
    \centering
    \includegraphics[width=1\linewidth]{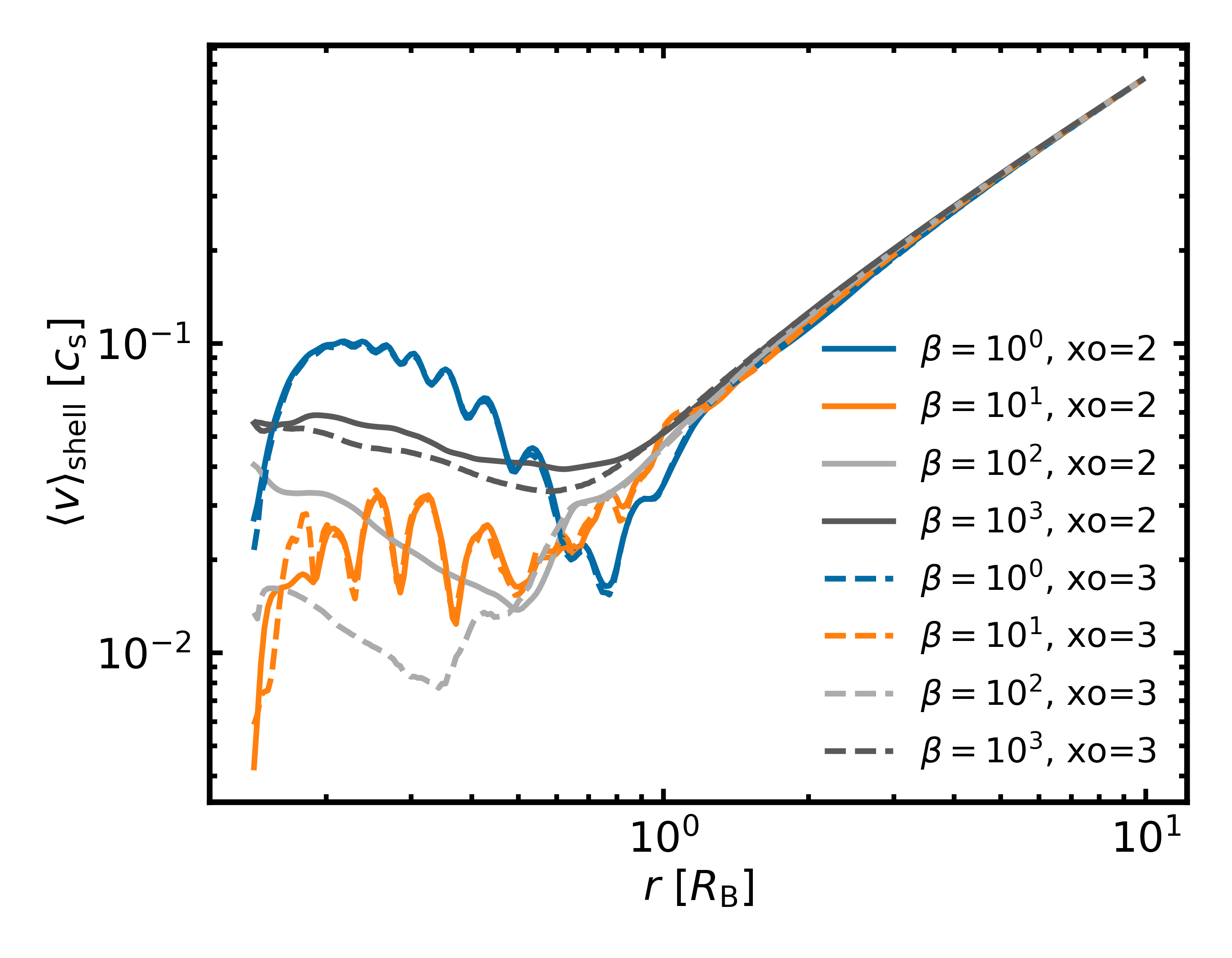}
    \caption{Shell-averaged mean flow speed for different $\beta$ and spatial reconstruction method. We set $m=0.1$, $N_r=256$, $r_{\rm in}=10\,R_{\rm p}$, and $\Phi_{\mathrm{p}} = -GM_{\mathrm{p}}/r \times f_{\mathrm{sm}} \times f_{\mathrm{inj}}$.}
    \label{fig:appendix flowspeed}
\end{figure}

\section{Convergence tests}\label{sec:Convergence tests}
In this section, we investigate how the simulation results depend on numerical settings: the resolution in the radial direction, $N_r$, the size of the inner boundary, $r_{\rm in}$, the spatial reconstruction method (denoted as \texttt{xoder=2} or \texttt{3}), and the form of the gravitational potential. Throughout the main text, we adopt $N_r = 256,\,r_{\rm in}=10\,R_{\rm p}$, \texttt{xorder=2}, and the gravitational potential defined as $\Phi_{\mathrm{p}} = -GM_{\mathrm{p}}/r \times f_{\mathrm{sm}} \times f_{\mathrm{inj}}$ (Eq. \ref{eq:Fung potential}).

Figure \ref{fig:appendix sph_avg_temp} shows the shell-averaged temperature profiles for a fully convective envelope (with $m = 0.1$ and $\beta = 10^3$). All curves are in excellent agreement, except for the one using a gravitational potential of the form $\Phi_{\mathrm{p}} = -GM_{\mathrm{p}}/\sqrt{r^2 + r_{\mathrm{sm}}^2}$ (the black dashed-dotted line). \cite{zhu2021global} argued that high spatial resolution is required to model the envelope accurately, resolving the Bondi radius with approximately 76 grids in the radial direction. In our numerical setup, this corresponds to $N_r > 160$. As shown in \Figref{fig:appendix sph_avg_temp}, all runs with $N_r = 168,\, 196$, and 256 yield nearly identical results, confirming sufficient resolution.

Using the potential of $\Phi_{\mathrm{p}} = -GM_{\mathrm{p}}/\sqrt{r^2 + r_{\mathrm{sm}}^2}$ causes non-zero gravitational force at the inner boundary, leading to a temperature jump at $r_{\mathrm{in}}$, as well as a sharp spike in the temperature gradient (\Figref{fig:appendix grad}). A similar temperature gradient jump is observed when adopting \texttt{xorder=3}. Additionally, although higher-order PPM method (\texttt{xorder=3}) is generally considered more suitable for maintaining hydrostatic balance than lower-order PLM method \citep[\texttt{xorder=2};][]{stone2020athena++,zhu2021global}, shell-averaged quantities exhibit fluctuations in the \texttt{xorder=3} case not present in the \texttt{xorder=2} case. The origin of the fluctuation is unclear, but it is certainly not caused by the averaging method. Nevertheless, across a wide radial range excluding the inner boundary, the results for \texttt{xorder=2} and \texttt{3} agree well despite the fluctuations (Figs. \ref{fig:appendix grad} and \ref{fig:appendix flowspeed}). Therefore, we adopt \texttt{xorder=2} as the default spatial reconstruction method in this study.

\def\thesection{B}
\setcounter{equation}{0}
\def\theequation{B.\arabic{equation}}
\setcounter{figure}{0}
\def\thefigure{B.\arabic{figure}}

\section{Fitting formulae for $r_{\rm RCB}^{\rm fit}$ and $r_{\rm RRB}^{\rm fit}$}\label{sec:Appendix fitting formulae}
\editsec{
We introduce the fitting formulae for the boundary between the radiative and convective (RCB) \editthird{layers}, and the recycling and radiative (RRB) layers. We used the \texttt{scipy.optimize.curve\_fit} library of python to constrain the fitting coefficients. We assumed that the numerical results, $r_{\rm RCB}^{\rm num}$ (Eq. \ref{eq:rcb num}), can be fitted by the following sigmoid curve:
\begin{align}
    r_{\rm RCB}^{\rm fit}=r_{\rm in}+a_{\rm fit}\bigg[1+\tanh\bigg\{b_{\rm fit}\big(\log_{10}\beta - c_{\rm fit}\big)\bigg\}\bigg]\quad(\beta>1).\label{eq:rcb fit}
\end{align}
Here $a_{\rm fit}=0.042,\,b_{\rm fit}=5.26$, and $c_{\rm fit}=2.09$ are the fitting coefficients. Equation \ref{eq:rcb fit} approaches $r_{\rm RCB}^{\rm fit}\rightarrow r_{\rm in}$\editthird{,} due to the limitation of the computational domain. We then assumed that the numerically-calculated RRB can be fitted by,
\begin{align}
    &r_{\rm RRB}^{\rm fit}=
    \begin{cases}
        \displaystyle
        \min\Bigg(R_{\rm atm}^{\rm KK24},\,R_{\rm atm}^{\rm KK24}\times\beta^{d_{\rm fit}}\Bigg)\quad(\beta\leq1),\\
        \displaystyle
        R_{\rm atm}^{\rm KK24}\Bigg(1+\frac{r_{\rm in}}{R_{\rm B}}-\frac{r_{\rm RCB}^{\rm fit}}{R_{\rm B}}\Bigg)\quad(\beta>1),
    \end{cases}
\end{align}
with $d_{\rm fit}=0.22$ being the fitting coefficient \editthird{and $R_{\rm atm}^{\rm KK24}$ determined as in \Equref{eq:Ratm,fit}}. 
}

\def\thesection{C}
\setcounter{equation}{0}
\def\theequation{C.\arabic{equation}}
\setcounter{figure}{0}
\def\thefigure{C.\arabic{figure}}

\section{Characteristic convective flow speed}\label{sec:Characteristic convective flow speed}
\begin{figure}[htbp]
    \centering
    \includegraphics[width=1\linewidth]{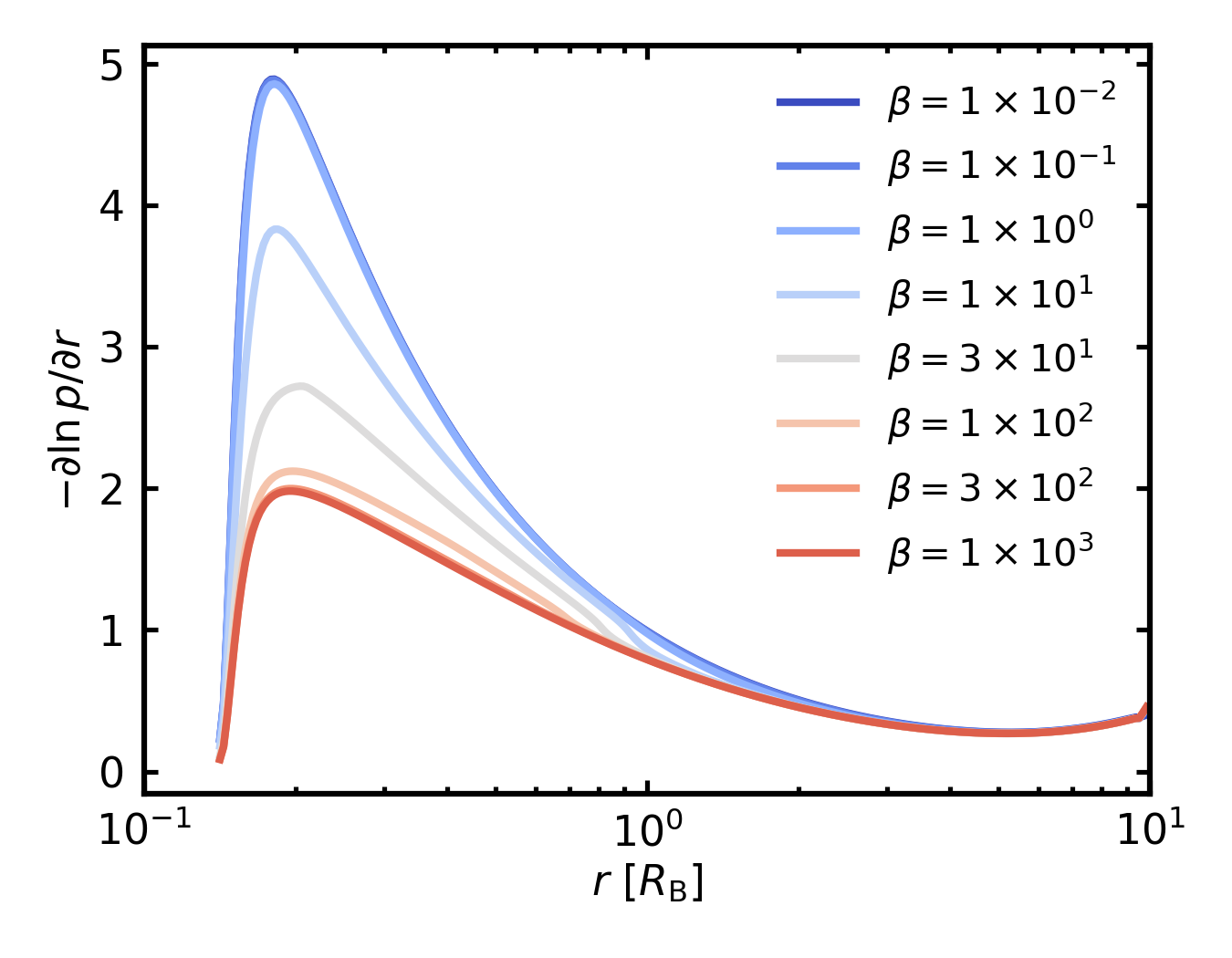}
    \caption{Shell-averaged pressure gradient for different $\beta$. We set $m=0.1$ and $\dot{M}=10\,M_\oplus$/Myr.}
    \label{fig:pgrad}
\end{figure}

Here we consider the energy is transported by convection, deriving the characteristic convective flow speed based on the mixing length theory \citep[e.g.,][]{kippenhahn1990stellar,ali2020limits}. The convective energy flux is given by
\begin{align}
    F_{\rm con}=\rho c_{\rm P}v_{\rm con}\Delta T,\label{eq:Fcon}
\end{align}
where $\Delta T$ is the temperature difference between a convective element and the surrounding. Then the luminosity is given by
\begin{align}
    L_{\rm acc}&=4\pi r^2F_{\rm con}.\label{eq:Lcon}
\end{align}
The convective element is subject to a buoyancy force per unit mass, 
\begin{align}
    F_{\rm b}&=-\frac{\Delta \rho}{\rho}\Bigg(\frac{GM_{\rm P}}{r^2}\Bigg)\approx-\frac{\Delta T}{T}\Bigg(\frac{1}{\rho}\frac{\partial p}{\partial r}\Bigg),\label{eq:Fb}
\end{align}
where $\Delta\rho$ is the density difference between the convective element and the surrounding. In the above equation, we used a hydrostatic equation and assumed a pressure equilibrium. We consider that the convective element moves by a certain distance from $0$ to $l_{\rm m}$ due to the work done by the buoyancy force. Then the average length can be set to $l_{\rm m}/2$. We assume that the half of the work by the buoyancy force is equal to the kinetic energy, 
\begin{align}
    F_{\rm b}\times\frac{l_{\rm m}}{2}\times\frac{1}{2}=\frac{v_{\rm con}^2}{2}.\label{eq:Energy conservation}
\end{align}
From Eqs. (\ref{eq:Fcon})--(\ref{eq:Energy conservation}) together with the Meyer's relation, we have
\begin{align}
    v_{\rm con}&=\Bigg[\frac{1}{8\pi r^2}\frac{\gamma-1}{\gamma}\frac{1}{R_{\rm g}T}\Bigg(-\frac{1}{\rho}\frac{\partial p}{\partial r}\Bigg)\frac{l_{\rm m}L_{\rm acc}}{\rho}\Bigg]^{1/3},\\
    &=\Bigg[\frac{1}{8\pi r^3}\frac{\gamma-1}{\gamma}\Bigg(-\frac{\partial \ln p}{\partial \ln r}\Bigg)\frac{l_{\rm m}L_{\rm acc}}{\rho}\Bigg]^{1/3},\\
    &\approx\Bigg(\frac{1}{8\pi r^3}\frac{\gamma-1}{\gamma}\frac{l_{\rm m}L_{\rm acc}}{\rho}\Bigg)^{1/3}    
\end{align}
where $R_{\rm g}$ is the gas constant. In the above equation, we assumed $p=\rho R_{\rm g}T$ and $-\partial\ln p/\partial \ln r\approx\mathcal{O}(1)$ (\Figref{fig:pgrad}).

\def\thesection{D}
\setcounter{equation}{0}
\def\theequation{D.\arabic{equation}}
\setcounter{figure}{0}
\def\thefigure{D.\arabic{figure}}

\section{\edit{Photospheric radius}}\label{sec:Photospheric radius}
\begin{figure}
    \centering
    \includegraphics[width=1\linewidth]{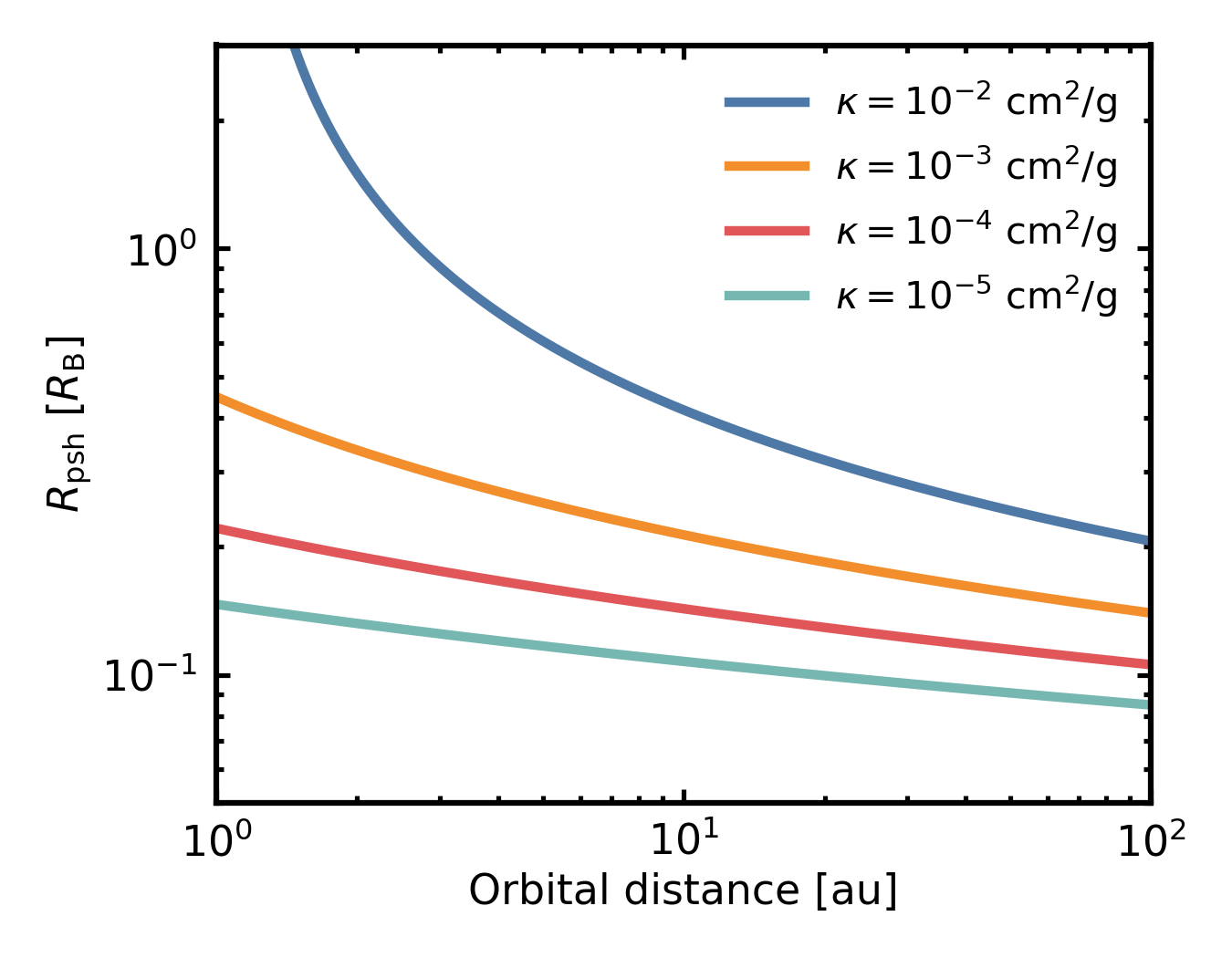}
    \caption{Photospheric radius for different opacities as a function of the orbital distance.}
    \label{fig:Rpsh}
\end{figure}

The optical depth of the envelope depends on the opacity, and in particular, the deep part of the envelope, inside the photosphere, is always optically thick \citep{rafikov2006atmospheres}. The radius of the photosphere is given by 
\begin{align}
    R_{\rm psh}\approx\frac{R_{\rm B}}{\ln(\lambda_{\rm mfp}/R_{\rm B})},\label{eq:Rpsh}
\end{align}
where $\lambda_{\rm mfp}=1/(\kappa\rho_0)$ is the mean free path. Equation \ref{eq:Rpsh} is plotted in \Figref{fig:Rpsh} for different opacities. 
Our simulations are conducted over a domain that extends beyond $r\gtrsim0.1\,R_{\rm B}$, and therefore, the assumption of optically thin conditions is valid when $\kappa\lesssim10^{-5}\text{--}10^{-4}\,\text{cm}^2\text{/g}$ or at distances outside of $\gtrsim$10 au.

\def\thesection{E}
\setcounter{equation}{0}
\def\theequation{E.\arabic{equation}}
\setcounter{figure}{0}
\def\thefigure{E.\arabic{figure}}

\section{\edit{Cooling time estimation}}\label{sec:Cooling time estimation}

The cooling time of the background gas is determined by two components \citep{miranda2020gaps,ziampras2023modelling}: surface cooling (radiative cooling from the disk surface) and in-plane cooling (radial thermal diffusion through the disk midplane). This is given by \citep{ziampras2023modelling}:
\begin{align}
    &\beta_{\rm surf}\equiv\frac{\Sigma_0\,c_{\rm V}}{2\sigma_{\rm SB}T_0^3}\tau_{\rm eff}\,\Omega,\quad\tau_{\rm eff}=\frac{3\tau}{8}+\frac{\sqrt{3}}{4}+\frac{1}{4\tau},\quad\tau=\frac{\kappa\Sigma_0}{2},\\
    &\beta_{\rm mid}\equiv\frac{3\kappa\rho_0c_{\rm V}}{16\sigma_{\rm SB}T_0^3}\Bigg(H^2+\frac{l_{\rm rad}^2}{3}\Bigg)\,\Omega,\quad l_{\rm rad}=\frac{1}{\kappa\rho_0},\\
    &\beta_{\rm disk}\equiv\frac{\beta_{\rm surf}}{1+\beta_{\rm surf}/\beta_{\rm mid}}.
    \label{eq:beta disk}
\end{align}
Here $\Sigma_0$ is the gas surface density, $T_0$ is the midplane temperature, $\rho_0=\Sigma_0/(\sqrt{2\pi}H)$ is the gas density, $c_{\rm V}=k_{\rm B}/(\mu m_{\rm H}(\gamma-1))$ is the specific heat capacity, $\kappa$ is the opacity, and $\sigma_{\rm SB}$ is the Stefan-Boltzmann constant. Equation \ref{eq:beta disk} accounts for both optically thick and optically thin regimes.

We consider a two-layer envelope model \citep{rafikov2006atmospheres,piso2014minimum}, assuming the envelope consists of an inner convective layer and an outer radiative layer. The cooling time, namely the time it takes to cool the envelope until the radiative-convective boundary (RCB) reaches a given pressure depth, is then computed as the ratio of the total energy contained between the core surface and the RCB to the luminosity emerging from the RCB \cite{piso2014minimum}:
\begin{align}
    \beta_{\rm RCB}\approx 4\pi \frac{p_{\rm RCB}}{L_{\rm RCB}\sqrt{R_{\rm p}}}\Bigg(\frac{\nabla_{\rm ad}}{\chi}R_{\rm B}\Bigg)^{1/\nabla_{\rm ad}}\Omega.\label{eq:beta RCB}
\end{align}
Here, 
\begin{align}
    &p_{\rm RCB}=\psi\, p_0e^{R_{\rm B}/R_{\rm RCB}},\\
    &L_{\rm RCB}\approx L_0\frac{p_0}{p_{\rm RCB}},\quad L_0=\frac{64\pi GM_{\rm p}\sigma_{\rm SB}T_0^{4}}{3\kappa p_0}\nabla_{\rm ad}\chi^{4},
\end{align}
are the pressure at the RCB, $p_0$ is the pressure at the midplane, and the emergent luminosity from the RCB with $\psi=0.556$ and $\chi=1.53$ being the constants \citep{piso2014minimum}. 
Here we adopt a conservative approach to maintain consistency with the two-layer envelope model in \cite{piso2014minimum}. However, we confirmed that using $L_{\rm acc}$ in  \Equref{eq:beta RCB} instead of $L_{\rm RCB}$ has little impact on the following results.

We define the envelope cooling time by:
\begin{align}
    \beta_{\rm env}=\max(\beta_{\rm disk},\,\beta_{\rm RCB}),\label{eq:beta env}
\end{align}
For a fixed planetary mass, orbital distance and a specific disk model, $\beta_{\rm env}$ is determined as a function of opacity and the RCB location, as shown in \Figref{fig:beta_env}. We use a passively irradiated disk model to compute the physical quantities at the midplane \citep{oka2011evolution},
\begin{align}
    &\Sigma_0=2.7\times10^3\,\text{g/cm}^2\,\bigg(\frac{a_{\rm p}}{1\,\text{au}}\bigg)^\xi,\quad T_0=150\,\text{K}\,\bigg(\frac{a_{\rm p}}{1\,\text{au}}\bigg)^{\zeta}.\label{eq:oka disk model}
\end{align}
\edit{We choose here a purely irradiated disk with} $\xi=-15/14$ and $\zeta=-3/7$.

\end{appendix}
\end{document}